\documentclass{article}

\usepackage[final]{neurips_2022}

\usepackage[utf8]{inputenc} % allow utf-8 input
\usepackage[T1]{fontenc}    % use 8-bit T1 fonts
\usepackage{hyperref}       % hyperlinks
\usepackage{url}            % simple URL typesetting
\usepackage{booktabs}       % professional-quality tables
\usepackage{amsfonts}       % blackboard math symbols
\usepackage{nicefrac}       % compact symbols for 1/2, etc.
\usepackage{microtype}      % microtypography
\usepackage{xcolor}         % colors

\usepackage{subfigure}

\usepackage{multirow}
\usepackage{tabularx}
\usepackage{adjustbox}
\usepackage{tikz}

\usepackage{wrapfig}
\usepackage{mathtools}
\usepackage[normalem]{ulem}
\usepackage{tabularx}
\usepackage{threeparttable}
\usepackage{amsmath,bm}
\usepackage{changepage}
\usepackage{amssymb}
\usepackage{mathtools}
\newtheorem{exmp}{Example}

\title{APG: Adaptive Parameter Generation Network for Click-Through Rate Prediction}

\author{
 Bencheng Yan$^{*}$, Pengjie Wang \thanks{$*$ These authors contributed equally to this work and are co-first authors.}, Kai Zhang, Feng Li, Hongbo Deng, Jian Xu, Bo Zheng \thanks{$\dagger$ Corresponding author}\\
{Alibaba Group} \\
{China} \\
\texttt{\{bencheng.ybc,pengjie.wpj,victorlanger.zk,}\\
\texttt{adam.lf,dhb167148,xiyu.xj,bozheng\}@alibaba-inc.com}\\
 }
\begin{document}

\maketitle

\begin{abstract}
In many web applications, deep learning-based CTR prediction models (deep CTR models for short) are widely adopted. 
Traditional deep CTR models learn patterns in a static manner, i.e., the network parameters are the same across all the instances. 
However, such a manner can hardly characterize each of the instances which may have different underlying distributions. 
It actually limits the representation power of deep CTR models, leading to sub-optimal results. 
In this paper, we propose an efficient, effective, and universal module, named as Adaptive Parameter Generation network (APG), which can dynamically generate parameters for deep CTR models on-the-fly based on different instances. 
Extensive experimental evaluation results show that APG can be applied to a variety of deep CTR models and significantly improve their performance. 
Meanwhile, APG can reduce the time cost by 38.7\% and memory usage by 96.6\% compared to a regular deep CTR model.
We have deployed APG in the industrial sponsored search system and achieved 3\% CTR gain and 1\% RPM gain respectively.
% \footnote{Note this is a significant improvement in the sponsored search system of Taobao, the largest e-commerce platform in China.}.
% where the gross revenue from sponsored search has been more than 100 billion per year.}.
\end{abstract}

% \vspace{-2em}

\section{Introduction}
\label{sec:Introduction}
% \vspace{-1em}
\begin{wrapfigure}{r}{0.5\columnwidth}
\vspace{-1em}
  \centering
    % \subfigure[The distribution of CTR from different user groups]
    % \subfigure{
    \includegraphics[trim=10 0 0 10, clip,width= .22\columnwidth]{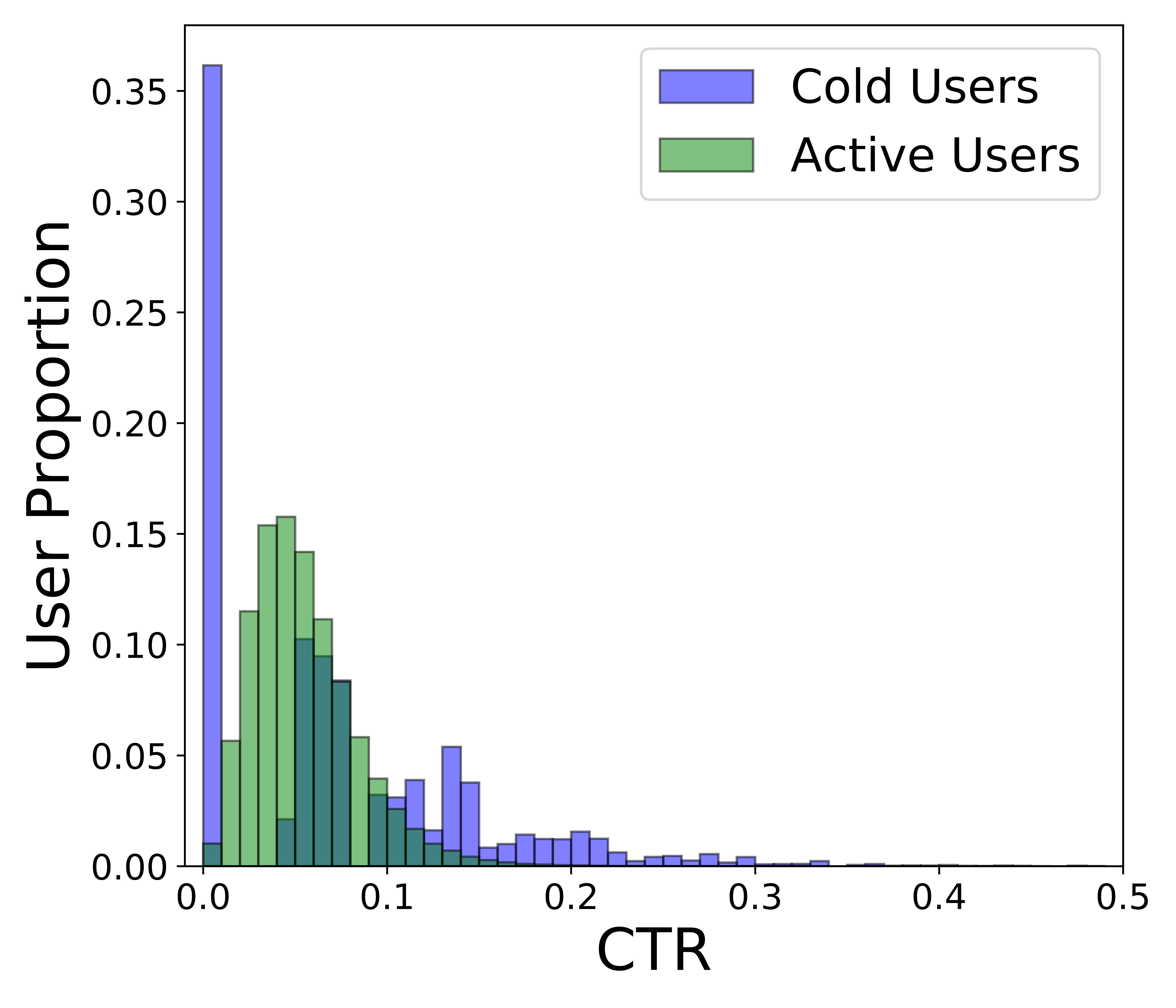}
    % \label{fig:subfigure2}
  % }
      % \subfigure[The distribution of age level from different user groups]
    % \subfigure{
    \includegraphics[trim=10 0 0 10, clip,width= .22\columnwidth]{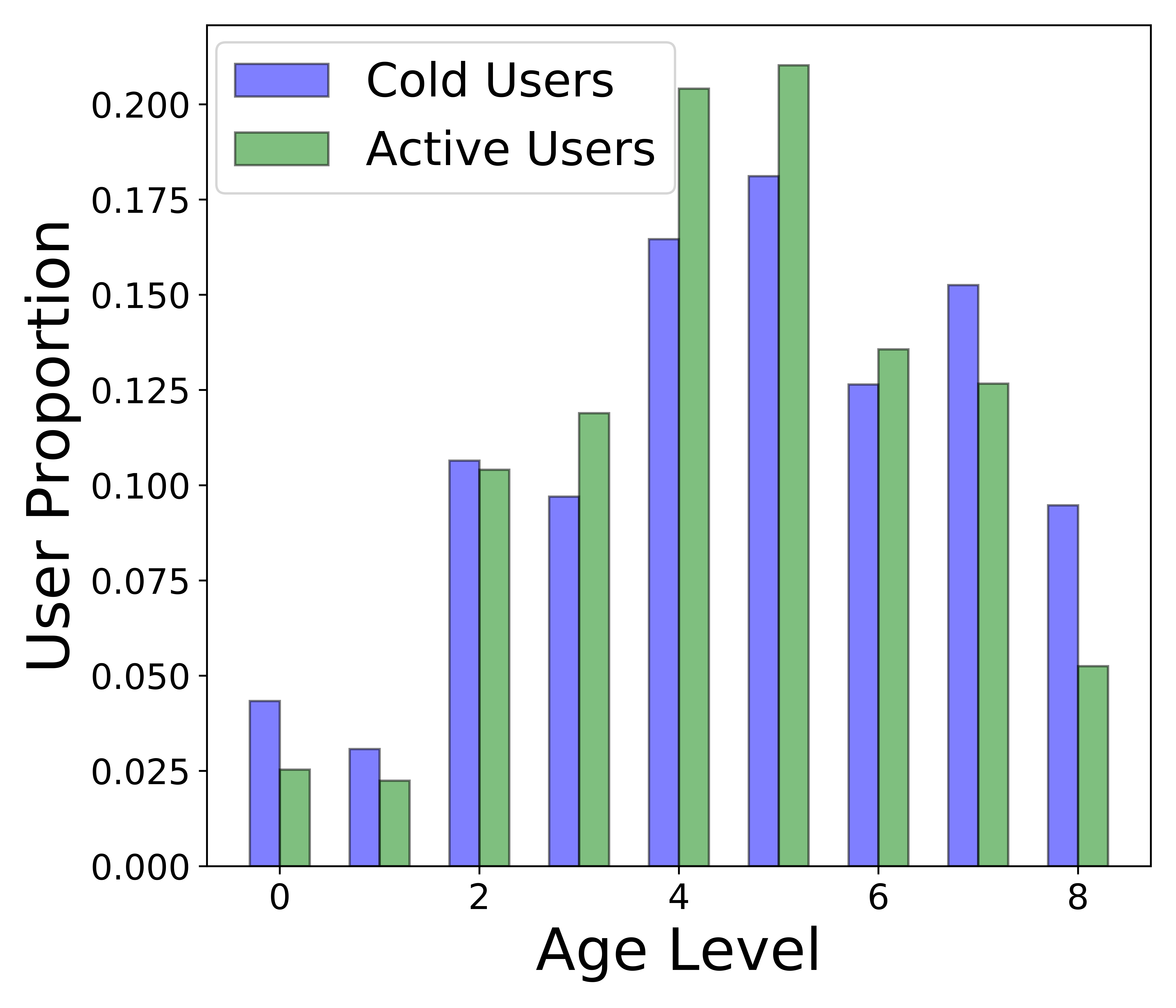}
    % \label{fig:subfigure2}
  % }
\vspace{-1em}
\caption{An example of feature distribution from different users (i.e., active vs. cold users). \textbf{Left:} the CTR distributions are varied from different user groups and a custom pattern should be considered. \textbf{Right:} A common pattern is welcomed to model the similar age distributions from different groups.}
\vspace{-1em}
\label{figure:An example of data distribution from different users (i.e., active vs. cold users)}
\end{wrapfigure}

Recently, deep CTR models have achieved great success in various web applications such as recommender systems, web search, and online advertising \cite{cheng2016wide,guo2017deepfm,wang2017deep,joglekar2020neural}. 
Formally, a regular deep CTR model can be expressed as $y_i=\mathcal{F}_{\Theta}(\bm{x}_i)$ where $\bm{x}_i,y_i$ are the input features and the predicted CTR of the instance $i$ respectively, $\Theta$ is the parameter, and $\mathcal{F}$ is usually implemented as a neural network.

Improving the performance of deep CTR models has been a very hot topic in the research and industrial areas.
Existing works can be broadly divided into two categories:
(1) Focusing on $\bm{x}_i$, more and more elaborated information (e.g., user behavior features \cite{zhou2018deepdin,pi2020search}, multimodal information \cite{chen2019personalized,he2021click}, knowledge graph \cite{zhao2019intentgc,wang2019knowledge}, etc) is introduced to enrich feature space (i.e., $\bm{x}_i$);
% and helps model better understand instances;
% (e.g., rich user behavior features are introduce in MIMN\cite{pi2019practice} and SIM \cite{pi2020search}).
(2) Focusing on $\mathcal{F}$, advanced architectures (including feature interaction modeling \cite{cheng2016wide,guo2017deepfm,wang2017deep}, automated architecture search \cite{joglekar2020neural,song2020towards} and so on) are designed to improve the model performance.
% (i.e, $\mathcal{F}$) are proposed to better capture the relationships among features, leading to better performance (e.g., WDL\cite{cheng2016wide}, DeepFM\cite{guo2017deepfm}, DCN\cite{wang2017deep}, etc).

% \begin{figure}[t]
% % \vspace{-1em}
%   \centering
%     \subfigure[The distribution of CTR from different user groups]{
%     \includegraphics[trim=10 0 0 10, clip,width= .22\textwidth]{User_CTR.png}
%     \label{fig:subfigure2}
%   }
%       \subfigure[The distribution of age level from different user groups]{
%     \includegraphics[trim=10 0 0 10, clip,width= .22\textwidth]{User_age.png}
%     \label{fig:subfigure2}
%   }

% \vspace{-1.5em}
% \caption{An example of data distribution from different users (i.e., active vs. cold users). (a) A custom pattern should be designed to capture the CTR distribution of different user groups. (b) A common pattern is welcomed to contribute each groups on age distribution modeling.}
% \vspace{-1em}
% \label{figure:An example of data distribution from different users (i.e., active vs. cold users)}
% \end{figure}

However, few works focus on the improvement of the model parameters $\Theta$, especially for the weight matrix $\bm{W} \in \mathbb{R}^{N \times M}$ used in hidden layers of deep CTR models \footnote{For simplicity, in this paper, we mainly discuss the weight matrix $W$.
Our method can also be easily applied to the parameters of other modules (e.g., transformer, attention network, etc) in deep CTR models since most of them can be regarded as a variety of MLP with a set of weight matrices \cite{zhang2021deep}.
% Let other parameters (e.g., used in embedding layer, batch norm layer, etc) be the future work.
}.
It is another orthogonal aspect for the performance improvement.
Actually, most of the existing works simply adopt a static manner, i.e., all the instances share the same parameters $\bm{W}$.
We argue that such a manner is sub-optimal for pattern learning and limits the representation power of deep CTR models.
On the one hand, although the common patterns among instances can be captured by the shared parameters $\bm{W}$,  it is not friendly to custom pattern modeling.
Specifically, taking the industrial sponsored search system as an example, the feature distribution can be varied from different users (e.g., active vs. cold users), different categories (e.g., clothing vs. medicine), and so on (see Figure \ref{figure:An example of data distribution from different users (i.e., active vs. cold users)} (a) as an example). 
Simply applying the same parameters across all the instances can hardly capture the characteristic of each instance from different distributions.
On the other hand, the learned common pattern may not be suitable for each of the instances.
For example, the shared parameters tend to be dominated by high-frequency features and may give a misleading decision for the long-tailed instances.
% In other words, the model capacity or representation power is limited and underestimated by such a static manner.
This leads us to the following question: 
\textit{Do we really need the same and shared parameters for all instances?}

Ideally,  besides modeling the common pattern, the parameters should be more adaptive and can be dynamically changed for different instances to capture custom patterns at the same time.
Then, the representation power (or model capacity) can be enhanced by the dynamically changed parameter space.
 % even if the weight matrix size is unchanged (e.g., $N \times M$).
% Then, such a custom parameter can easily capture the custom patterns of each instance from different distributions.
To achieve this goal, we design a new paradigm for CTR prediction.
 % and design input-aware parameters.
The key insight is to propose an Adaptive Parameter Generation network (APG) to dynamically generate parameters depending on different instances.
Firstly, we propose a basic version (Section \ref{sec:Basic Model}) of APG which can be expressed as $y_i=\mathcal{F}_{\mathcal{G}(\bm{z}_i)}(\bm{x}_i)$ where $\mathcal{G}$ refers a neural network (e.g., MLP) and generates the adaptive parameters $\bm{W}_i \in \mathbb{R}^{N \times M}$ by the input-aware condition $\bm{z}_i \in \mathbb{R}^{D}$.
% which is various from different instances $i$. 
However, the basic model suffers from two problems:
(1) inefficient in time and memory.
Directly generating the weight matrix $\bm{W}_i$ of a deep CTR model needs $\mathcal{O}\textit{(NMD)}$ cost in computation and memory storage, which is $D$ times the cost in a regular deep CTR model and is costly especially for a web-scale application where  $N,M$ are usually set as a large value (e.g., $\textit{N=M=1,000}$).
The empirical results (Section \ref{sec:Efficiency Evaluation}) also show the basic model needs an extra 111$\times$ training time and 31$\times$ memory usage.
(2) sub-effective in pattern learning.
% It is hardly for the basic model to 
The parameter generation process is totally dependent on the condition $\bm{z}_i$, which may only capture the custom patterns and ignore the common patterns which contribute to understand users' behaviors (see Figure \ref{figure:An example of data distribution from different users (i.e., active vs. cold users)} (b)), leading to sub-effective pattern learning.

Then, we extend the above basic model to an efficient and effective version of APG:
 \textbf{(1) For the efficiency,} 
motivated by the low-rank methods \cite{li2018measuring,aghajanyan2020intrinsic} which show that the weight matrix resides on a low intrinsic dimension, we parameterize the target weight matrix $\bm{W}_i$ as the production of three low-rank matrices $\bm{U}_i\bm{S}_i\bm{V}_i$ where $\bm{U}_i \in \mathbb{R}^{N{\times}K}, \bm{V}_i \in \mathbb{R}^{K{\times}M},\bm{S}_i \in \mathbb{R}^{K{\times}K}$ and \textit{$K \ll min(M,N)$}.
 % (Section \ref{sec:Low-rank parameterization}).
In addition, the decomposed feed-forwarding 
% (Section \ref{sec:Decomposed Feed-forwarding})
is proposed to avoid the heavy computation of the weight matrix $\bm{W}_i$ reconstruction.
 % to significantly reduce the computation and storage cost from $\mathcal{O}\textit{(NMD)$ to $\mathcal{O}\textit{((NK+MK)D)$ and keep high performance at the same time  (see Section \ref{sec:Low-rank parameterization}). 
% Then, to further reduce the complexity of the generation process and efficiently model the common patterns, 
Then we further take the center matrix $\bm{S}_i$ as the specific parameters which are dynamically generated to capture the custom patterns and the rest two matrices $\bm{U,V}$ as the shared parameters which are randomly initialized  and shared across instances to capture common patterns.
 % (Section \ref{sec:Parameter sharing}).
In this way, the complexity of generating the specific parameters can be easily controlled by setting a small $K$. 
 % resulting in the computation and storage complexity of $\mathcal{O}\textit{(KKD)$ which is free from the large value $N,M$  (see Section \ref{sec:Parameter sharing}).
% Finally, thanks to the low-rank parameterization, we propose a Decomposed Feed-Forwarding which reduces the time cost of in a deep CTR model from $\mathcal{O}\textit{(NM)$ to $\mathcal{O}\textit{((N+M)K+KKD)$  (see Section \ref{sec:Parameter Factorization}).
% Besides, with the further proposed Decomposed Feed-Forwarding, the total time cost of a deep CTR model with APG per layer is reduced to $\mathcal{O}\textit{((N+M)K+KKD)$  compared with the cost  $\mathcal{O}\textit{(NM)$ in a regular deep CTR model (see Section \ref{sec:Model Complexity});
As a result, APG achieves $\mathcal{O}\textit{(KKD+NK+MK+KK)}$ time complexity and $\mathcal{O}\textit{(KKD+NK+MK)}$ memory complexity compared to that both have $\mathcal{O}\textit{(NM)}$ in a regular deep CTR model.
We empirically find APG can speed up the training time by 38.7\% and reduce memory usage by 96.6\% relative to a regular deep CTR model (Section \ref{sec:Efficiency Evaluation}).
 \textbf{(2) For the effectiveness,} 
% besides introducing two kinds of parameters to model the custom and common patterns (mentioned above), we further propose some strategies to improve these two pattern learning respectively.
% (a) For the shared parameters, we extend $\bm{U}$ and $\bm{V}$ to an over parameterization version which enriches the model capacity without any additional memory and time cost during inference (see Section \ref{sec:Over Parameterization}).
% (b) For the specific parameters, we borrow the idea of conditional computation \cite{bengio2015conditional,cho2014exponentially} and propose three kinds of condition strategies for $\bm{z}_i$ to meet various requirements in real applications.
% Furthermore, we also find the semantic relationships among different generated $\bm{S}_i$ which implicitly model the common information (see Section \ref{sec:Visualization}).
% besides introducing two kinds of parameters to model the custom and common patterns (mentioned above), % we further propose some strategies to improve these two pattern learning respectively.
apart from the natural effectiveness of APG in custom pattern learning, to model the common pattern, the shared weights $\bm{U}$ and $\bm{V}$ are considered in the adaptive parameter generation.
Then we further extend $\bm{U}$ and $\bm{V}$ to an over parameterization version which enriches the model capacity without any additional memory and time cost during inference.
 % (Section \ref{sec:Over Parameterization}).
Besides, we also find there exists inherent similarity between different generated $\bm{S}_i$ which implicitly model the common information (Section \ref{sec:Visualization}).

In summary, the contributions of this paper are presented as follows:
% \begin{itemize}
% \item
(1) We propose a new learning paradigm in deep CTR models where the model parameters are input-aware and dynamically generated to boost the representation power. 
It is orthogonal to many prior methods and is a universal module that can be easily applied in most existing deep CTR models.
% \item 
(2) We present APG which generates the adaptive parameters in an efficient and effective way, and theoretically analysis the computation and memory complexity.
% On the one hand, low-rank parameterization  decomposed feed-forwarding is proposed to significantly reduce the memory and computation cost.
% On the other hand, the condition design, the shared parameters, and over parameterization are presented to capture the custom and common patterns effectively.  
% On the other hand, the design of condition and shared parameters is presented to capture the custom and common patterns effectively.
% \item  
(3) Extensive experimental evaluation results demonstrate that the proposed method is a universal module and can improve the performance of most of the existing deep CTR models. 
we also provide a systematic evaluation of APG in terms of training time and memory consumption.
Finally, we have developed APG in the industrial sponsored search system and achieved 3\% CTR gain and 1\% RPM gain respectively.
% \end{itemize}
% \begin{itemize}
% \item We propose a new learning paradigm in deep CTR models where the model parameters are input-aware and dynamically generated to boost the representation power. 

% \item We present APG, an efficient, powerful, and universal module, which designs data-aware condition $\bm{z}_i$ firstly, and then, uses this condition to generate adaptive parameters by Re-Parameterization.

% \item Extensive experimental evaluation results demonstrate that the proposed method outperforms strong baseline methods, and can also improve the performance of most of the existing deep CTR models. 
% Besides, we also develop APG on the industrial application, and the online results show the great advantages of the proposed module.
% \end{itemize}

% \vspace{-1em}
\section{Related Work}
\label{sec:Related Work}
% \vspace{-1em}
% In this section, we introduce some works related to our papers, including standard deep CTR model and predefined-based methods.
% \subsection{Deep CTR Models}
\textbf{Deep CTR Models.}
A traditional CTR prediction method usually adopts a deep neural network to capture the complex relations between users and items.
Most of them focus on (1) introducing abundant useful informations \cite{zhou2018deepdin,pi2020search,chen2019personalized,he2021click,zhao2019intentgc,wang2019knowledge} to improve the model understanding,
(2) designing advanced architecture \cite{cheng2016wide,guo2017deepfm,wang2017deep,joglekar2020neural,song2020towards} to achieve better performance.
All of them adopt a static parameters manner which limits the model capacity of these methods, leading to sub-optimal performance

\textbf{Coarse-grained Parameter Allocating.}
There are some research areas that bring a coarse-grained parameter allocating strategy that may be related to our goals, including multi-domain learning \cite{sheng2021one,lee2019melu} and multi-task learning \cite{ma2018modeling,misra2016cross}.
Both of them maintain and allocate different parameters to different domains or tasks manually.
However, such a coarse-grained parameter allocating can hardly be extended to a fine-grained (e.g., user, item, or instances sensitive) manner.
It not only costs too much memory to maintain and store a large number of parameters when considering the fine-grained modeling but also is lack flexibility and generalization since the parameter allocation is manually pre-defined.
We also conduct experiments to compare this kinds of methods (Section \ref{sec:Comparison the performance with predefined-based methods}).

% \subsection{Dynamic Deep Neural Networks}
\textbf{Dynamic Deep Neural Networks.}
Our method is related to recent works of dynamic neural networks used in computer vision and natural language processing, in which the model parameters \cite{jia2016dynamic,wang2021adaptive,fan2018decouple,platanios2018contextual} and architecture \cite{liu2017learning,yuan2020s2dnas} can be dynamically changed.
In our paper, we take the first step to bring the idea about dynamic networks into deep CTR models and develop it in real applications where the efficiency in computation and memory is extremely needed and the common and custom pattern learning are also required.

% \vspace{-1em}
\section{Method}
% \vspace{-1em}
In this paper, we denote scalars, vectors and tensors with lower-case (or upper-case), bold lower-case, and bold upper-case letters, e.g., $n$ (or $N$), $\bm{x}$, $\bm{X}$, respectively.
% The framework of the proposed APG is depicted in Figure \ref{figure:The framework of APG.}.
% General speaking, APG contains two main processes including specific parameter generation and adaptive parameter generation.
% The former generates specific parameter by input-aware conditions. 
% The latter combines the specific parameters and shared parameters to generate the adaptive parameters used for deep CTR models.

\subsection{Basic Model}
\label{sec:Basic Model}
In this section, we introduce the basic version of APG. 
General speaking, the basic idea of APG is to dynamically generate parameters $\bm{W}_i$ by different condition $\bm{z}_i$, i.e., $\bm{W}_i = \mathcal{G}{(\bm{z}_i)}$
% \begin{align}
% \bm{W}_i = \mathcal{G}{(\bm{z}_i)}
% \end{align}
% \begin{align}
% \Theta_i = \mathcal{G}{(\bm{z}_i)}
% \end{align} 
where $\mathcal{G}$ refers to the adaptive parameter generation network.
Then the generated parameters are applied to the deep CTR models, i.e., $y_i=\mathcal{F}_{\mathcal{G}{(\bm{z}_i)}}(\bm{x}_i)$.
% \begin{align}
% y_i=\mathcal{F}_{\mathcal{G}{(\bm{z}_i)}}(\bm{x}_i)
% \end{align}
% Since the parameters of deep CTR models usually refers to the weight matrices of different layers or modules, in this paper, we focus on generating the weight matrix $\bm{W}_i$ of the model (Note other parameters can also be generated in the same way), i.e.,
% \begin{align}
% \bm{W}_i = \mathcal{G}{(\bm{z}_i)}
% \end{align}
% Then  we need to answer the following two questions:
% \begin{itemize}
% % \setlength{\itemindent}{-.2em}
%  \item \textbf{Q1}: What is the condition $\bm{z}_i$ for different instance $i$?
%  \item \textbf{Q2}: How to generate parameters efficiently and effectively?
% \end{itemize}
Next, we present (1) how to design the condition $\bm{z}_i$ for different instance $i$  and (2) how to implement $\mathcal{G}$.

\subsubsection{Condition Design}
\label{sec:Condition Design}
% parameter space design
% As a matter of fact, $\bm{z}_i$ takes an important role of APG since different $\bm{z}_i$ gives different prior knowledge on the parameter space design.
In this paper, we propose three kinds of strategies (including group-wise, mix-wise, and self-wise) to design different kinds of $\bm{z}_i$.

% \subsubsection*{\normalfont{\textbf{Group-wise}}}
\textbf{Group-wise.}
The group-wise strategy tries to generate different parameters depending on different instance groups.
The purpose is that, sometimes, the instances can be divided into different groups \cite{cao2010aworldwide} and instances in the same group may have similar patterns.
Thus $\bm{z}_i$ is the representation identifying different groups.
% Thus, designing the adaptive parameters for each group can help the model to better characterize the data and give a more suitable pattern for different groups.
An example can be found in Appendix \ref{sec:group-wise example}.

% \subsubsection{Mix-wise}
% \subsubsection*{\normalfont{\textbf{Mix-wise}}}
% \label{sec:Mix-wise}
\textbf{Mix-wise.}
To further enrich the expression power and flexibility of the adaptive parameters, a mix-wise strategy is designed to take multiple conditions into consideration.
% Compared with group-wise strategy, co-wise strategy take multiple conditions into consideration and give a more complex prior knowledge to the parameters space design.
Specifically, given $k$ condition embeddings $\{\bm{z}^j_i \in \mathbb{R}^d | j\in \{0,1,..,k-1\}\}$ 
% (For simplicity, the embedding dimensions of different conditions are set the same) 
of the instance $i$, we propose two aggregation policies to consider different conditions at the same time:
% \subsubsection*{\normalfont{\textbf{Input Aggregation.}}}
(1) \textit{Input Aggregation.}
This policy firstly aggregates different condition embeddings and then feeds the aggregated embeddings into APG to obtain the mix-wise based parameters. 
The aggregation functions can be (but are not limited to): Concatenation, Mean, and Attention.
% (a) Concatenation function: $\bm{z}_i=concat(\bm{z}^j_i)$;
% (b) Mean function: $\bm{z}_i=1/k \sum_j \bm{z}^j_i$;
% (c) Attention function: $\bm{z}_i=\sum_j \alpha_j \bm{z}^j_i$ where $\alpha_j$ refers to the attention sore calculated by self-attention.
% $\alpha_j=\bm{W_{att}}\bm{z^j}_i/\sum_t \bm{W_{att}}\bm{z^t}_i$ and $\bm{W_{att}} \in \mathbb{R}^{1 \times d}$
(2) \textit{Output Aggregation}
% \subsubsection*{\normalfont{\textbf{Output Aggregation.}}}
This policy firstly feeds different $\bm{z}^j_i$ into different APG and obtains the corresponding parameters $\bm{W}^j_i$ respectively.
Then the aggregation is applied to these adaptive parameters $\bm{W}^j_i$. 
Similarly, the aggregation function includes Concatenation, Mean, and Attention.
Examples can be found in Appendix \ref{sec:mix-wise example}.

% \subsubsection{Self-wise}
% \subsubsection*{\normalfont{\textbf{Self-wise}}}
\textbf{Self-wise.}
The above two strategies need additional prior knowledge to generate parameters.
Self-wise strategy tries to use a simple and easily obtained knowledge (i.e., self-knowledge) to guide parameters generation, i.e., the network parameters are generated by their own input.
For example, for the $0$-th hidden layer of a deep CTR model, we can set $\bm{z}_i = \bm{x}_i$.
For $l$-th hidden layer, $\bm{z}_i = \bm{h^{l-1}}$ where $\bm{h^{l-1}}$ is the input of the $l$-th hidden layer.
% In this strategy, $\bm{z}_i \in \mathbb{R}^M$ where $M$ is the width of the current hidden layer.

\begin{figure*}[t]
\centering
\includegraphics[width = \textwidth]{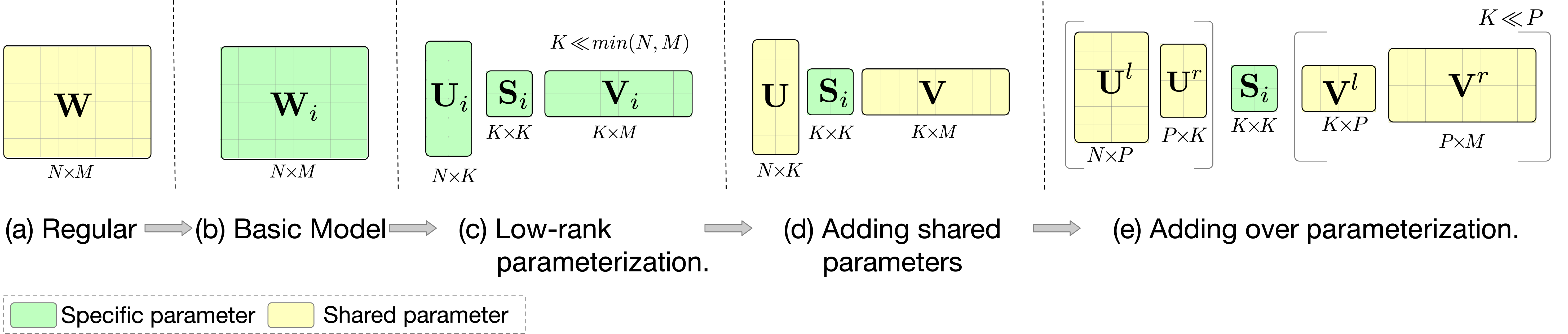}
\vspace{-0.5em}
\caption{The comparison of different versions of APG.
}
\vspace{-1em}
\label{figure:Re-Parameterization.}
\end{figure*}

\subsubsection{Parameters Generation}
After obtaining the conditions, we adopt a multilayer perceptron\footnote{In this paper, we take MLP as an example, and other implementations of $\mathcal{G}$ can also be considered.} to generate parameters depending on these conditions (see Figure \ref{figure:Re-Parameterization.} (b)), i.e.,
\begin{align}
\label{equ:raw parameters generate}
\bm{W}_i = reshape(MLP(\bm{z}_i))
\end{align} where $\bm{W}_i \in \mathbb{R}^{N \times M}$, $\bm{z}_i \in \mathbb{R}^{D}$ 
% \footnote{When adopting group-wise or mix-wise strategy, $D=d$. When adopting self-wise strategy, $D=M$}
, and the operation $reshape$ refers to reshaping the vectors produced by the MLP into a matrix form.
Then a deep CTR model with APG can be expressed as:
\begin{align}
\label{equ:raw deep ctr model with APG}
y_i = \sigma(\bm{W}_i\bm{x}_i)
\end{align}where $\sigma$ is the activation function.
Here we take a deep CTR model with an MLP layer as an example, and other deep CTR models can also be easily extended since most of them can be regarded as a variety of
MLP with a set of weight matrices \cite{zhang2021deep}.

\subsection{Effective and Efficient Adaptive Parameter Generation Network}
\label{sec:Effective and Efficient Adaptive Parameter Generation Network}
\begin{wrapfigure}{r}{0.4\columnwidth}
\centering
\includegraphics[trim=10 0 0 0, clip,width = .4\columnwidth]{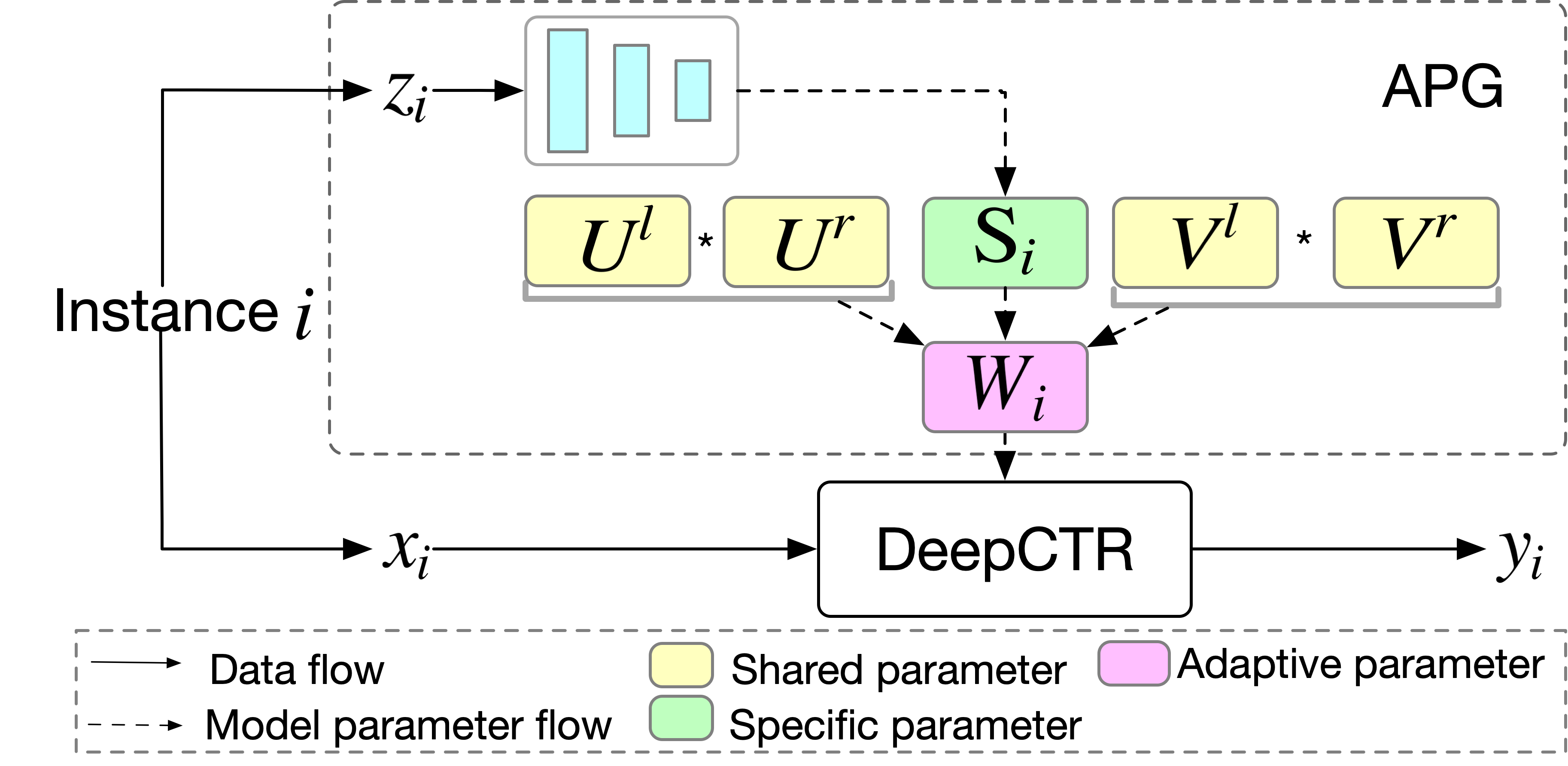}
\vspace{-1em}
\caption{The framework of APG. For simplicity, the process of decomposed feed-forwarding is omitted in this figure.
}
\vspace{-1em}
\label{figure:The framework of APG.}
\end{wrapfigure}
% The above basic model has two problems: 
% (1) \textbf{Inefficient.}
% Assuming that APG adopts a single perceptron layer in Eq \ref{equ:raw parameters generate},
% the generation needs $\mathcal{O}\textit{(NMD)}$ computation cost and $\mathcal{O}\textit{(NMD)}$ memory cost.
% Then the total computation complexity is $\mathcal{O}\textit{(NMD+NM)}$ per layer where $\textit{(NM)}$ refers to the cost of the feed-forwarding in Eq \ref{equ:raw deep ctr model with APG}.
% It means the computation (or memory costs) are $D$ times larger than that in a regular deep CTR model per layer (see Table \ref{table:The computation complexity of different versions of APG.}.
% Such a huge extra time and memory cost will definitely influence the model serving and may not be tolerated by real applications;
% (2) \textbf{Sub-effective.}
% In the basic model, $\bm{W}_i$ is totally dependent on the given condition $\bm{z}_i$, which may ignore the common pattern modeling.
% % Although it can characterize the custom patterns of different instances, the common patterns among different instances may be ignored.
% Actually, such a common pattern usually gives useful information for CTR prediction.
% To address the above two challenges, we propose four extensions including low-rank parameterization, decomposed feed-forwarding, parameter sharing, and over parameterization to parameterize the weight matrix in an efficient and effective way.
%  % to Parameter Factorization and Over Parameterization to reconstruct the parameter space (called Re-Parameterization).
As introduced in Section \ref{sec:Introduction}, the above basic model has two problems: (1) inefficient in time and memory and (2) sub-effective in pattern learning.
% Assuming that APG adopts a single perceptron layer in Eq \ref{equ:raw parameters generate},
% the generation needs $\mathcal{O}\textit{(NMD)}$ computation cost and $\mathcal{O}\textit{(NMD)}$ memory cost.
% Then the total computation complexity is $\mathcal{O}\textit{(NMD+NM)}$ per layer where $\textit{(NM)}$ refers to the cost of the feed-forwarding in Eq \ref{equ:raw deep ctr model with APG}.
% It means the computation (or memory costs) are $D$ times larger than that in a regular deep CTR model per layer (see Table \ref{table:The computation complexity of different versions of APG.}.
% Such a huge extra time and memory cost will definitely influence the model serving and may not be tolerated by real applications;
% (2) \textbf{Sub-effective.}
% In the basic model, $\bm{W}_i$ is totally dependent on the given condition $\bm{z}_i$, which may ignore the common pattern modeling.
% Although it can characterize the custom patterns of different instances, the common patterns among different instances may be ignored.
% Actually, such a common pattern usually gives useful information for CTR prediction.
To address the above two problems, we propose some extensions including low-rank parameterization, decomposed feed-forwarding, parameter sharing, and over parameterization to parameterize the weight matrix in an efficient and effective way.
 % to Parameter Factorization and Over Parameterization to reconstruct the parameter space (called Re-Parameterization).

% \subsubsection{Low-rank parameterization}
% \label{sec:Low-rank parameterization}
\textbf{Low-rank parameterization.}
Inspired by the recent success of the low-rank methods \cite{li2018measuring,aghajanyan2020intrinsic} which have demonstrated that strong performance can be achieved by optimizing a task in a low-rank subspace, we hypothesize that the adaptive parameters also have a low “intrinsic rank”.
To this end, we propose to parameterize the weight matrix $\bm{W}_i \in \mathbb{R}^{N \times M}$ as a low-rank matrix, which is the product of three sub-matrices $\bm{U}_i \in \mathbb{R}^{N \times K}, \bm{S}_i \in \mathbb{R}^{K \times K}, \bm{V}_i \in \mathbb{R}^{K \times M}$ and the rank $K \ll min(N,M)$ (see Figure \ref{figure:Re-Parameterization.} (c)).
Formally, the weight generation process can be expressed as:
\begin{align}
\label{equ:low-rank}
\bm{U}_i,\bm{S}_i,\bm{V}_i &= reshape(MLP(\bm{z}_i)) 
\end{align}
Intuitively, we can set a small value of $K$ to control the time and memory cost.
Meanwhile, ignoring the substantial storage and computation cost, $K$ can also be set to a higher value to enlarge the adaptive parameter space (see detailed discussion in Appendix \ref{sec:Evaluation of the hyper-parameters}).
Overall, through the low-rank parameterization, we can significantly reduce dimensionality in the parameter space, i.e., $K \ll min(N,M)$, to enable a more compact model.
% Specifically, replacing the generation of the large weight matrix $\bm{W}_i$ in Eq \ref{equ:raw parameters generate} with the three sub-matrices in Eq \ref{equ:low-rank}.
As a result, the computation complexity of the specific parameter generation process can be reduced to $\mathcal{O}\textit{((NK+MK+KK)D)}$ (see Table \ref{table:The computation complexity of different versions of APG.}).
% Since $K \ll min(N,M)$, the complexity cost is approximated to  $\mathcal{O}\textit{((NK+MK)D)}$.
The memory cost is also reduced to $\mathcal{O}\textit{((NK+MK+KK)D)}$ (see Table \ref{table:The computation complexity of different versions of APG.}).

% \begin{figure}[t]
%   \centering
%     \subfigure[An example of the decomposed feed-forwarding.]{
%     \includegraphics[width = .45\textwidth]{Decomposed_Feed-forwarding.png}
%     \label{figure:An example of the decomposed feed-forwarding..}
%   }
%       \subfigure[An example of over parameterization.]{
%     \includegraphics[width = .45\textwidth]{Over-param.png}
%     \label{figure:Over Parameterization.}
%   }
% \vspace{-1em}
% \caption{An example of data distribution from different users (i.e., active vs. cold users). (a) A custom pattern should be designed to capture the CTR distribution of different user groups. (b) A common pattern is welcomed to contribute each groups on age distribution modeling.}
% \vspace{-1em}
% \label{figure:An example of data distribution from different users (i.e., active vs. cold users)}
% \end{figure}

\begin{figure}
\centering
\begin{minipage}{.5\textwidth}
  \centering
  \includegraphics[width=.9\linewidth]{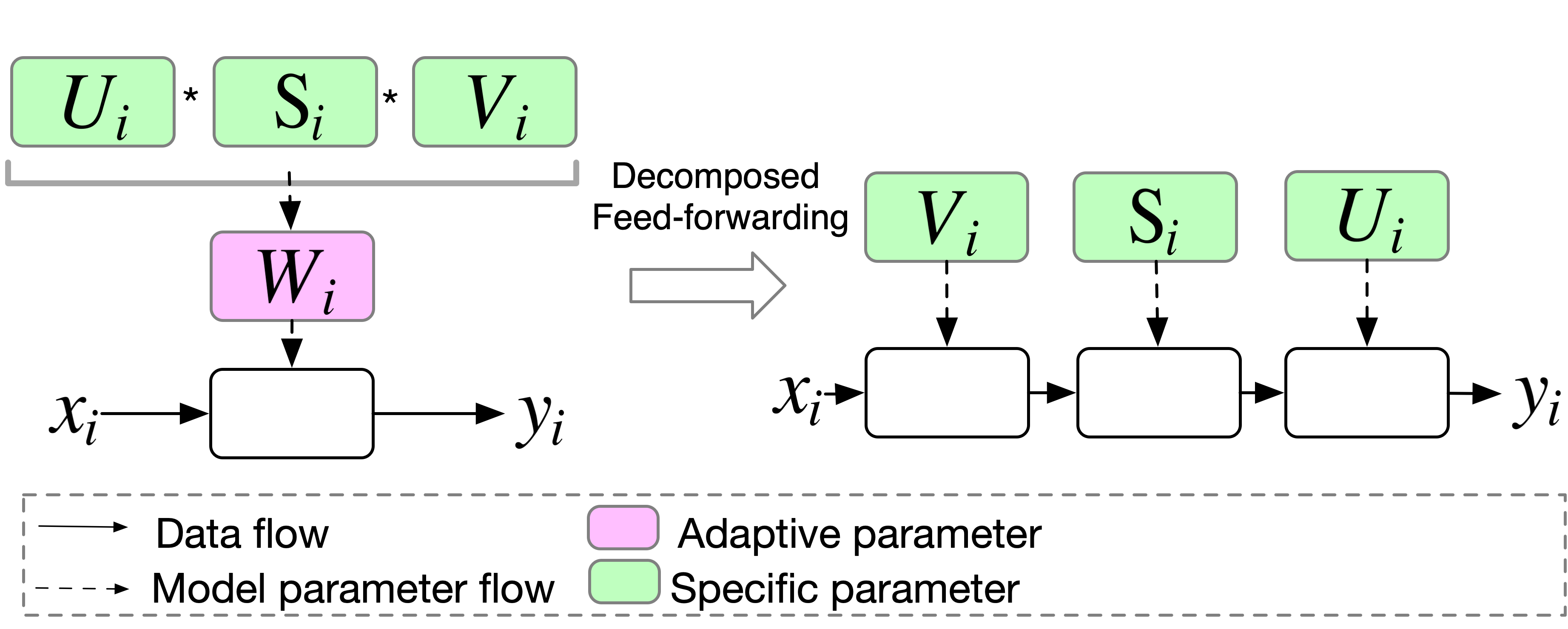}
  \caption{An example of the decomposed feed-forwarding.}
  \label{figure:An example of the decomposed feed-forwarding..}
\end{minipage}%
\begin{minipage}{.5\textwidth}
  \centering
  \includegraphics[width=.9\linewidth]{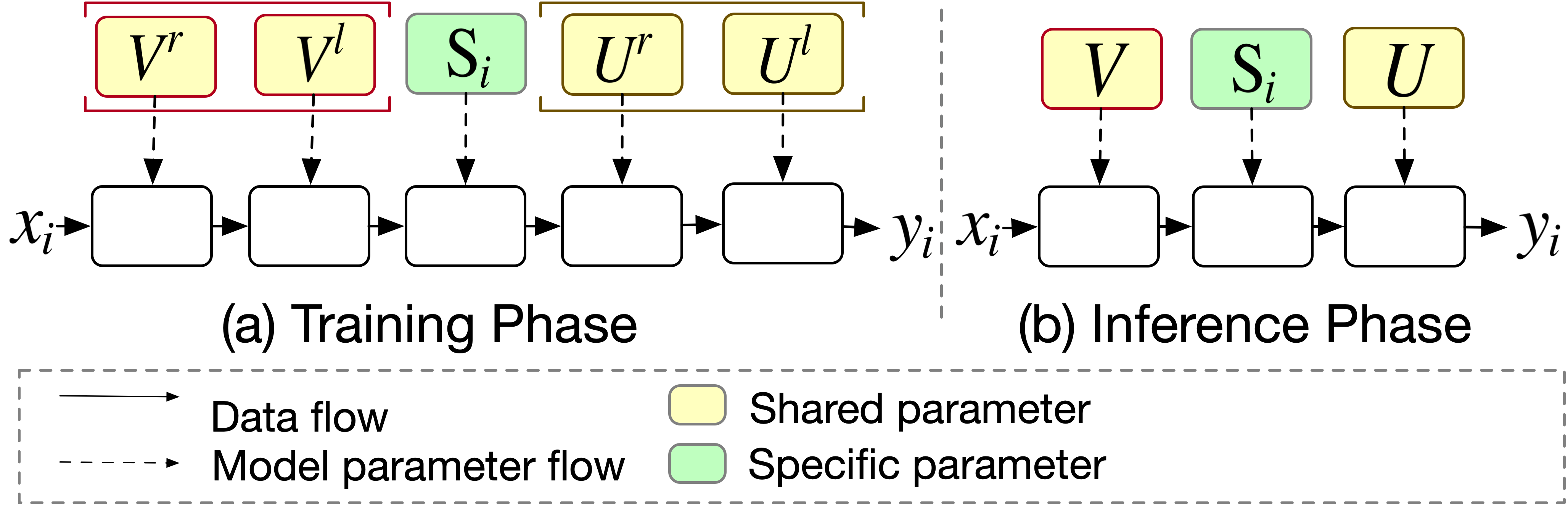}
  \caption{An example of over parameterization.}
  \label{figure:Over Parameterization.}
\end{minipage}
\vspace{-1em}
\end{figure}

% \subsubsection*{\normalfont{\textbf{Decomposed Feed-forwarding.}}}
% \subsubsection{Decomposed Feed-forwarding}
% \label{sec:Decomposed Feed-forwarding}
\textbf{Decomposed Feed-forwarding.}
After low-rank parameterization, the Eq \ref{equ:raw deep ctr model with APG} can be written as
\begin{align}
\label{equ:DFF}
y_i&=\sigma(\bm{W}_i\bm{x}_i) = \sigma((\bm{U}_i\bm{S}_i\bm{V}_i)x_i)=\sigma(\bm{U}_i(\bm{S}_i(\bm{V}_i \bm{x}_i)))
\end{align}
Here instead of directly reconstructing the weight matrix $\bm{W}_i$ by the sub-matrix production, we design a decomposed feed-forwarding and apply $\bm{x}_i$ to each sub-matrix sequentially (see Figure \ref{figure:An example of the decomposed feed-forwarding..}).
Such a design helps us avoid the heavy computation of the sub-matrix production which costs $\mathcal{O}\textit{(NKK+NMK)}$.
Actually, this design benefits from the low-rank parameterization which naturally supports the decomposed feed-forwarding.
Since the computation complexity in Eq \ref{equ:DFF} is $\mathcal{O}\textit{(NK+KK+MK)}$, 
% which is approximated to $\mathcal{O}\textit{(NK+MK)}$ due to $K \ll min(N,M)$, 
the total time cost of a deep CTR model with APG per layer is reduced to $\mathcal{O}\textit{((NK+MK+KK)(D+1))}$ (see Table \ref{table:The computation complexity of different versions of APG.}).

\textbf{Parameter sharing.}
% As described in Section \ref{sec:Introduction}, the common patterns also take an important role in CTR prediction.
In this section, we present our common pattern modeling strategy.
Thanks to decomposing the weight matrix into $\bm{U}_i, \bm{S}_i$, and $\bm{V}_i$, it allows us to be more flexible.
Consequently, we divide these three matrices into 
(1) \textit{specific parameters} that capture custom patterns from different instances;
(2) \textit{shared parameters} that are shared across instances to characterize the common patterns.
Specifically, we define $\bm{S}_i$ as the specific parameters since the matrix scale is totally controlled by $K$ and is more efficient for the generation process.
 $\bm{U}$ and $\bm{V}$ are regarded as the shared parameters (see Figure \ref{figure:Re-Parameterization.} (d)).
 Then we can rewrite Eq \ref{equ:low-rank} and \ref{equ:DFF} as:
 \begin{align}
\label{equ:svd}
\bm{S}_i &= reshape(MLP(\bm{z}_i)) \\
y_i &= \sigma(\bm{U}(\bm{S}_i(\bm{V} \bm{x}_i)))
\end{align}
Furthermore, such a design also contributes the efficiency.
Since the size of the generated specific parameter is reduced to $K \times K$, the computation complexity of the specific parameter generation can be further reduced to $\mathcal{O}\textit{(KKD)}$ in Eq \ref{equ:svd}, and the total time cost of a deep CTR model with APG is reduced to $\mathcal{O}\textit{(KKD+NK+MK+KK)}$ (see Table \ref{table:The computation complexity of different versions of APG.}).
Meanwhile the memory cost is reduced to $\mathcal{O}\textit{(KKD+NK+MK)}$ where $\textit{NK}$ and $\textit{MK}$ refers to the storage of the shared parameters $\bm{U}$ and $\bm{V}$ respectively (see Table \ref{table:The computation complexity of different versions of APG.}).
\textbf{Over Parameterization}
Compared with the shared weight matrix $\bm{W} \in \mathbb{R}^{N \times M}$ in a regular deep CTR model, $\bm{U}$ and $\bm{V}$ in APG can be hardly scaled to large matrices due to the efficiency constraint with $K \ll min(N,M)$, leading to a possible performance drop.
% This may cause insufficient learning in custom patterns.
To address this problem and further enlarge the model capacity, 
% inspired by ACNet \cite{ding2019acnet} which adds more parameters in a convolution kernel without extra time cost, we design an over parameterization to enrich the number of the common parameters.
we follow the idea about the over parameterization \cite{aghajanyan2020intrinsic,ding2019acnet} to enrich the number of the shared parameters (see Figure \ref{figure:Re-Parameterization.} (e)).
Specifically, we replace shared parameters in Eq 6 with two large matrices, i.e.,
\begin{align}
\label{equ:over-param}
\bm{U} = \bm{U}^l \bm{U}^r, \;
\bm{V} = \bm{V}^l \bm{V}^r
\end{align} 
where $\bm{U}^l \in \mathbb{R}^{N \times P}$,$\bm{U}^r \in \mathbb{R}^{P \times K}$,$\bm{V}^l \in \mathbb{R}^{K \times P}$,$\bm{V}^r \in \mathbb{R}^{P \times M}$ and $P \gg K$.
Although $\bm{U}$ (or $\bm{V}$) is exactly equal to $\bm{U}^l \bm{U}^r$ (or $\bm{V}^l \bm{V}^r$) at the mathematical perspective, replacing $\bm{V}$ with $\bm{V}^l \bm{V}^r$ can contribute into two folds:
(1) Since $P \gg K$, more shared parameters are introduced to enlarge the model capacity \cite{ding2019acnet,ding2021diverse};
(2) The form as the multiple matrix production can result in an implicit regularization and thus enhance generalization \cite{aghajanyan2020intrinsic}.

% \begin{figure}[t]
% \centering
% \includegraphics[width = .45\textwidth]{Over-param.png}
% \caption{An example of over parameterization.
% }
% % \vspace{-1.5em}
% \label{figure:Over Parameterization.}
% \end{figure}

% \textbf{No Additional Inference Latency and Memory Cost.}
% \subsubsection*{\normalfont{\textbf{No Additional Inference Latency and Memory Cost.}}}
% The model can benefit from abundant learnable parameters
\textit{No Additional Inference Latency and Memory Cost.}
During the training phase, we can set $P$ to a large value to enlarge the model representation power.
Remarkably, when $P > max(N,M)$, the shared parameter space can be large than that in a regular deep CTR model.
During the inference phase, we explicitly pre-compute and store $\bm{V},\bm{U}$, and use these two matrices for inference (see Figure \ref{figure:Over Parameterization.}).
% equivalently transformed these two matrices into a single matrix by matrix products.
% In this way, we can introduce rebudent there is no extra time cost in inference
Critically, this guarantees that we can introduce abundantly shared parameters without any additional latency and memory cost during the inference phase.

\begin{table*}[t] \footnotesize
\caption{The computation and memory complexity per layer of different versions of APG. 
SPG refers to the time or memory cost in the process of specific parameter generation.
R-Wi is the time cost to reconstruct $\bm{W}_i$.
FF refers to the time cost in the feed-forwarding process of a regular deep CTR model.
SPS refers to the memory cost to store the shared parameters.
The total computation cost is the sum of SPG, R-Wi, and FF.
The total memory cost is the sum of SPG and SPS.
Since over parameterization dose not introduce any additional time cost, it is not depicted here.}
\resizebox{\columnwidth}{!}{
\begin{tabular}{l|cccc|ccc}
\toprule
 % & \multicolumn{4}{c}{Computation}  
 \multirow{2}{*}{\textbf{Versions}} & \multicolumn{4}{c|}{\textbf{Computation complexity per layer}}                              & \multicolumn{3}{c}{\textbf{Memory complexity per layer}}   \\  \cmidrule{2-8} 
% {}                        & \textbf{Cost}  & \textbf{$\Delta$}   & \textbf{Cost}  & \textbf{$\Delta$}   & \textbf{Cost}  & \textbf{$\Delta$}  & \textbf{Cost}  & \textbf{$\Delta$} & \textbf{Cost}  & \textbf{$\Delta$}     \\ \midrule
% \multicolumn{11}{c}{\textbf{Time / Epoch (s)}} \\ \midrule
  &SPG  &  R-Wi & FF &  Total  Cost& SPG & SPS &Total Cost\\ \midrule
 $\bm{Wx}_i$&      -         &          -     &  $\mathcal{O}\textit{(NM)}$ &  $\mathcal{O}\textit{(NM)}$       &  -         &        $\mathcal{O}\textit{(NM)}$        &    $\mathcal{O}\textit{(NM)}$    \\ \midrule
 $\bm{W}_i\bm{x}_i$&   $\mathcal{O}\textit{(NMD)}$  &          -     &  $\mathcal{O}\textit{(NM)}$ &   $\mathcal{O}\textit{(NMD+NM)}$     &       $\mathcal{O}\textit{(NMD)}$          &       -        &   $\mathcal{O}\textit{(NMD)}$    \\ \midrule
 $(\bm{U}_i\bm{S}_i\bm{V}_i) \bm{x}_i$&   $\mathcal{O}\textit{((NK+MK+KK)D)}$   &    $\mathcal{O}\textit{(NKK+NMK)}$  & $\mathcal{O}\textit{(NM)}$ &$\mathcal{O}\textit{((NK+MK+KK)D+NKK+NMK+NM)}$  &    $\mathcal{O}\textit{((NK+MK+KK)D)}$    &          -     &       $\mathcal{O}\textit{((NK+MK+KK)D)}$             \\ \midrule
 $\bm{U}_i(\bm{S}_i(\bm{V}_i \bm{x}_i))$&      $\mathcal{O}\textit{((NK+MK+KK)D)}$         &       -        &     $\mathcal{O}\textit{(NK+MK+KK)}$              &   $\mathcal{O}\textit{((NK+MK+KK)(D+1))}$ &    $\mathcal{O}\textit{((NK+MK+KK)D)}$            &        -       &   $\mathcal{O}\textit{((NK+MK+KK)D)}$     \\ \midrule
 $ \bm{U}(\bm{S}_i(\bm{V} \bm{x}_i))$&     $\mathcal{O}\textit{(KKD)}$          &      -         &     $\mathcal{O}\textit{(NK+MK+KK)}$                &   $\mathcal{O}\textit{(KKD+NK+MK+KK)}$  &        $\mathcal{O}\textit{(KKD)}$       &      $\mathcal{O}\textit{(NK+MK)}$         &    $\mathcal{O}\textit{(KKD+NK+MK)}$   \\ \bottomrule
\end{tabular}
}
\vspace{-2em}
\label{table:The computation complexity of different versions of APG.}
\end{table*}

\vspace{-1em}
\subsection{Complexity}
\label{sec:Model Complexity}
\vspace{-1em}
% 不同wise，不同 aggregation的复杂度 
In this section, we detailed analyze the proposed model complexity including the memory and computation complexity during the inference phase.
For analysis, the parameters generation network is implemented as a single perceptron layer, and the per layer costs in a regular deep CTR model and an adaptive deep CTR model are compared.
Summarization can be found in Table \ref{table:The computation complexity of different versions of APG.}.
% \noindent \textbf{Number of Parameters}
% % Here we analyze the number of learnable parameters
% For a standard deep CTR model, the number of parameters is $\mathcal{O}\textit{(NM)$ (i.e., the weight matrix) per layer.
% For APG, to obtain custom parameters $\bm{S}_i$ in Eq \ref{equ:svd}, we need a parameters generation network which cost $\mathcal{O}\textit{(KKD)$.

\textbf{Memory Complexity.}
% \subsubsection*{\normalfont{\textbf{Memory Complexity}}}
For a regular deep CTR model, the memory cost is $\mathcal{O}\textit{(NM)}$ per layer, i.e., storing the shared weight matrix.
For APG, the memory cost comes from two parts: 
(1) The memory cost of generating $\bm{S}_i$ in Eq \ref{equ:svd} is $\mathcal{O}\textit{(KKD)}$;
(2) The shared parameters $\bm{U},\bm{V}$ cost  $\mathcal{O}\textit{((N + M)K)}$ during the inference phase.
Then the total memory complexity of APG is $\mathcal{O}\textit{(KKD+NK+MK)}$ per layer.

\textbf{Computation Complexity.}
% \subsubsection*{\normalfont{\textbf{Computation Complexity}}}
A regular deep CTR model needs $\mathcal{O}\textit{(NM)}$ per layer for the feed-forward computation.
APG needs $\mathcal{O}\textit{(KKD)}$ in Eq \ref{equ:svd} to calculate the specific parameters.
Meanwhile, the feed-forwarding computation of a deep CTR model is $\mathcal{O}\textit{(NK+KK+MK)}$ by the decomposed feed-forwarding in Eq 8.
% As for condition generation, for group-wise and self-wise strategies, the condition embedding can be directly obtained from a deep CTR model without any extra time cost.
% For mix-wise strategy, 
In total, the computation complexity is $\mathcal{O}\textit{(KKD+NK+KK+MK)}$.
% Since $K \ll min(N,M)$, the computation cost is nearly $\mathcal{O}\textit{(KKD+NK+MK)}$.

In summary, APG has $\mathcal{O}\textit{(KKD+NK+MK)}$ in memory cost and $\mathcal{O}\textit{(KKD+NK+KK+MK)}$ in computation cost.
Since $K \ll min(N,M)$ and $D$ is usually set smaller than $N,M$,
% \footnote{The width (i.e., $M$ and $N$) of the hidden layer in a deep CTR model is usually scaled to the sum of dimensionality (e.g., $nd$) of all (e.g., $n$) feature embeddings. 
% Thus when adopting the group-wise or mix-wise strategy, the condition embedding dimensionality $D=d$  is usually much smaller than $N$ and $M$. 
% When adopting a self-wise strategy, $D=M$.}, 
the memory and computation cost of APG can even be much smaller than that (i.e., $\mathcal{O}\textit{(NM)}$) in a regular deep CTR model.
The experimental results in Section \ref{sec:Efficiency Evaluation} also show the efficiency of APG.

\section{Experiments}
\vspace{-1em}
\subsection{Experimental Settings}
\label{sec:overall Experimental Settings}
% \vspace{-1em}
The detailed settings including datasets, baselines, and training details are presented in Appendix \ref{sec:The detailed experimental setting}.

% \subsubsection*{\normalfont{\textbf{Datasets}}}
\textbf{Datasets.}
Four real-world datasets are used including \textbf{Amazon}, \textbf{MovieLens}, \textbf{IAAC}, and \textbf{IndusData}.
The first three are public datasets and the last is an industrial dataset.

% \subsubsection*{\normalfont{\textbf{Baselines}}}
\textbf{Baselines.}
Here, we compare our method with two kinds of methods
(1) \emph{Existing CTR prediction methods} include WDL\cite{cheng2016wide}, PNN\cite{qu2016product}, FIBINET\cite{huang2019fibinet}, DIFM\cite{lu2020dual}, DeepFM\cite{guo2017deepfm}, DCN\cite{wang2017deep}, and AutoInt\cite{song2019autoint};
(2) \emph{Coarse-grained parameter allocating methods.} multi-task learning: MMoE \cite{ma2018modeling} and multi-domain learning: Star \cite{sheng2021one}.
% Note due to the limited space, we present the results with the coarse-grained parameter allocating methods in Appendix \ref{sec:Comparison the performance with predefined-based methods}.

% \subsubsection*{\normalfont{\textbf{Training Details.}}}
\textbf{Training Details.}
See Appendix \ref{sec:Training Details} for the detailed introduction.

\subsection{Performance Evaluation with Existing Deep CTR Models}
\label{sec:Evaluation the performance with standard deep CTR models}
% \subsubsection{Results on public datasets}
% \subsubsection*{\normalfont{\textbf{Results on public datasets.}}}

\textbf{Results on public datasets.}
We first apply APG on CTR prediction tasks on public datasets. 
% i.e., models predict the probability that a user clicks the recommended item which is the main task in various scenarios like recommendation systems and so on.
Since APG is a universal module and can be applied to most of the existing deep CTR models.
% As introduced in Section \ref{sec:model}, APG is a general framework which can be applied to most of existing deep CTR models as long as the parameters of these deep CTR models are generated by APG (i.e., Eq \ref{equ:matrix-multiply} and \ref{equ:bias-multiply}).
Hence, to evaluate the performance of APG, we apply APG to various existing deep CTR models, and report the results of the original model (denoted as Base) and the model with APG.
Here, AUC (\%) \cite{fawcett2006introduction} score is reported as the metric \footnote{Note 0.1\% absolute AUC gain is regarded as significant for the CTR task \cite{zhou2018deepdin,autoint,cheng2016wide}.}.
The results are shown in Table \ref{tabel:Click-Through Rate (CTR) Prediction Task}.
We can find that with the help of APG, all of the methods achieve a significant improvement on all datasets.
% Detailedly, for the 
For example, the gains of DCN is 0.33\% $\sim$ 1.61\% (other methods also can obtain similar improvement).
It demonstrates that (1) giving adaptive parameters for models can enrich the parameter space and learn more useful patterns for different instances;
(2) the proposed APG is a universal framework that can boost the performance of many other methods.
Such nice property encourages APG to be applied to various scenarios and various methods.

\textbf{Results on industrial application.} 
We also develop APG in the industrial sponsored search system, and achieve 0.2\% AUC gain on the industrial dataset, 3\% CTR gain and 1\% RPM (Revenue Per Mile) gain during online A/B test.
Detailed analysis are presented in Appendix \ref{sec:Evaluation on the industrial application}.

\begin{table*}[t] \tiny
\caption{The AUC (\%) results of Click-Through Rate (CTR) prediction on different datasets. Note Base refers to the original results of the corresponding methods and Base+APG refers to the results with the help of APG.
Ave is the average results across all cases.
$\Delta$ refers the improvement of Base+APG compared to Base.}
\centering
\resizebox{\columnwidth}{!}{
\begin{tabular}{l|l|c|c|c|c|c|c|c|c}
\toprule
\textbf{Data}  &     \textbf{Method}     &\textbf{WDL}  & \textbf{PNN}   & \textbf{FIBINET} & \textbf{DIFM} & \textbf{DeepFM} & \textbf{DCN}   & \textbf{AutoInt} & \textbf{Ave} \\ 
\midrule
\multirow{3}{*}{MovieLens} & Base       & 79.21 & 79.5  & 79.78   & 79.84 & 79.3   & 79.29 & 79.36   & 79.46\\ 
% \cline{2-9} 
                           & Base+APG & \textbf{79.73} & \textbf{79.67} & \textbf{79.82}   & \textbf{79.94} & \textbf{79.60}  & \textbf{79.62}  & \textbf{79.64}  &\textbf{79.70} \\ 
                           & $\Delta$       & +0.39 & +0.17 & +0.04 & +0.10 &+0.30 &+0.33&+0.28&+0.24 \\
                           \midrule
\multirow{3}{*}{Amazon}    & Base       & 69.15 & 69.16 & 68.88   & 69.17 & 69.1   & 68.98 & 68.96  & 69.06 \\ 
% \cline{2-9} 
                           & Base+APG & \textbf{69.43} & \textbf{69.37} & \textbf{69.19}    & \textbf{69.23} & \textbf{69.43}  & \textbf{69.42} & \textbf{69.38}  & \textbf{69.34} \\ 
                           & $\Delta$       & +0.22 & +0.21 &+0.31&+0.06&+0.33&+0.44&+0.42&0.28 \\
                           \midrule
\multirow{3}{*}{IAAC}      & Base       & 65.17 & 65.3  & 65.15   & 65.76 & 65.64  & 64.78 & 64.99   &65.26\\ 
% \cline{2-9} 
                           & Base+APG & \textbf{65.94} & \textbf{65.87} & \textbf{66.15}   & \textbf{66.42} & \textbf{66.17}  & \textbf{66.39} & \textbf{66.21} &\textbf{66.15}  \\ 
                           & $\Delta$       & +0.77 &+0.57 & +1.0 &+0.66&+0.53&+1.61&+1.22&+0.91 \\
\bottomrule
\end{tabular}
}
\label{tabel:Click-Through Rate (CTR) Prediction Task}
\end{table*}

\begin{table}[t] \footnotesize
\begin{minipage}{.55\linewidth}
\caption{The settings of different APG versions. 
% Note LP refers to the low-rank parameterization. 
% $|\Theta_i|$ and $|\Theta|$ refer to the specific and shared parameter size in a deep CTR model respectively.
% DFF refers to the decomposed feed-forwarding.
% We report AUC (\%) as the performance metric, second/epoch as the time cost, and megabytes as the memory cost, and the corresponding ratio compared with Base is also presented.
% Besides, in this part, to avoid the influence of over parameterization, all of the versions of APG do not adopt over parameterization.
}
% \begin{adjustwidth}{-1.1cm}{}
\centering
\resizebox{\columnwidth}{!}{
\begin{tabular}{l|l|l}
\toprule
% \multirow{2}{*}{} & \multicolumn{5}{c|}{\textbf{Setting}}    & \multirow{2}{*}{\textbf{MovieLens}}  & \multirow{2}{*}{\textbf{Amazon}} & \multirow{2}{*}{\textbf{IAAC}} \\ \cmidrule{2-6} 
                  & \textbf{Version}         & \textbf{Annotation}   \\ \midrule
Base              & \scalebox{0.7}{$\bm{W}\bm{x}_i$}  & WDL \cite{cheng2016wide} as the baseline and the backbone \\ \midrule
v1                &\scalebox{0.7}{$\bm{W}_i\bm{x}_i$}              & Basic model   (Section \ref{sec:Basic Model})    \\ \midrule
v2                & \scalebox{0.7}{$(\bm{U}_i\bm{S}_i\bm{V}_i)\bm{x}_i$}  & + Low-rank parameterization    \\ \midrule
v3                & \scalebox{0.7}{$\bm{U}_i(\bm{S}_i(\bm{V}_i \bm{x}_i))$} & + Decomposed feed-forwarding      \\ \midrule
v4                & \scalebox{0.7}{$ \bm{U}(\bm{S}_i(\bm{V} \bm{x}_i))$}  & + Parameter sharing \\ \midrule
% v5                & \scalebox{0.7}{$ \bm{U}(\bm{W}_i(\bm{V} \bm{x}_i))$}                   & $\checkmark$  & \scalebox{0.7}{$NM$}    & \scalebox{0.7}{$NN+MM$}      & $\checkmark$                   & {{79.58}}     & {69.40}   & {66.04}         \\ \bottomrule
v5                & \scalebox{0.7}{$ (\bm{U}^l\bm{U}^r)(\bm{S}_i((\bm{V}^l\bm{V}^r)x_i))$}   & + Over parameterization      \\ \bottomrule
\end{tabular}
% \end{adjustwidth}
}
\label{table:The settings of different APG versions. }
\end{minipage}
% \end{table}
% \begin{table}[t] \footnotesize
\begin{minipage}{.45\linewidth}
\caption{The AUC results of the evaluation of different versions of APG. 
$\Delta$ refers the difference relative to Base.
% Note LP refers to the low-rank parameterization. 
% $|\Theta_i|$ and $|\Theta|$ refer to the specific and shared parameter size in a deep CTR model respectively.
% DFF refers to the decomposed feed-forwarding.
% We report AUC (\%) as the performance metric, second/epoch as the time cost, and megabytes as the memory cost, and the corresponding ratio compared with Base is also presented.
% Besides, in this part, to avoid the influence of over parameterization, all of the versions of APG do not adopt over parameterization.
}
% \begin{adjustwidth}{-1.1cm}{}
\centering
\resizebox{\columnwidth}{!}{
\begin{tabular}{l|c|c|c|c|c}
\toprule
% \multirow{2}{*}{} & \multicolumn{5}{c|}{\textbf{Setting}}    & \multirow{2}{*}{\textbf{MovieLens}}  & \multirow{2}{*}{\textbf{Amazon}} & \multirow{2}{*}{\textbf{IAAC}} \\ \cmidrule{2-6} 
                   & \textbf{MovieLens}& \textbf{Amazon}&      \textbf{IAAC}            & \textbf{Ave(AUC)} & \textbf{Ave($\Delta$)}            \\ \midrule
Base            & {79.21}      & {69.15}         & {65.17}  & 71.18&  --- \\ \midrule
v1                 & {79.51}      & {69.33}          & {65.52} &71.45  & +0.27                             \\ \midrule
% v2                & \scalebox{0.7}{$(\bm{U}_i\bm{S}_i\bm{V}_i)\bm{x}_i$}      & {79.56}      & {69.34}  & {65.67}      \\ \midrule
v2                 & {79.49}       & {69.24}   & {65.61}  & 71.45& +0.27  \\ \midrule
v4                 & {79.61}      & {69.36}    & {65.78}   & 71.58 & +0.40 \\ \midrule
% v5                & \scalebox{0.7}{$ \bm{U}(\bm{W}_i(\bm{V} \bm{x}_i))$}                   & $\checkmark$  & \scalebox{0.7}{$NM$}    & \scalebox{0.7}{$NN+MM$}      & $\checkmark$                   & {{79.58}}     & {69.40}   & {66.04}         \\ \bottomrule
v5                 & {{79.73}}     & {69.43}   & {65.94}   & 71.70 &   +0.52   \\ \bottomrule
\end{tabular}
}
% \end{adjustwidth}
\label{table:The AUC results of the evaluation about different versions of APG. }
\end{minipage}
\end{table}

\subsection{Effectiveness Evaluation}
\label{sec:Effectiveness Evaluation}
% \vspace{-1em}
In this section, we implement various versions of APG (see Table \ref{table:The settings of different APG versions. }) and conduct experiments to detailedly evaluate the effectiveness of APG.
The AUC results are reported in Table \ref{table:The AUC results of the evaluation about different versions of APG. }.
Note since the design of decomposed feed-forwarding does not influence the AUC performance of APG, we do not compare version v3 in this section.
% It is discussed in Section \ref{sec:Efficiency Evaluation} for the efficiency evaluation.

\textbf{The impact of the basic model.}
Compared with the base, v1 achieves significant improvements on AUC results overall datasets.
It demonstrates the effectiveness to introduce specific parameters to give a custom understanding of different instances.

\textbf{The impact of the low-rank parameterization.}
The purpose of low-rank parameterization is to reduce the computation and memory cost and keep high performance at the same time.
Considering the effectiveness, v1 and v2 do not provide much performance difference. 
Both of them achieve high performance and perform much better than Base.
More importantly, by low-rank parameterization, we can generate adaptive parameters in a more efficient way (see Section \ref{sec:Efficiency Evaluation} for details).

\textbf{The impact of the parameter sharing.}
% As introduced in Section \ref{sec:Introduction}, common patterns play an important role in CTR prediction.
In APG, we introduce the shared parameters to characterize common patterns.
% Here, we detailedly analyze the influence of the shared parameters.
Comparing the performance of versions v2, and v4 can further improve the performance by introducing the shared parameters, demonstrating the effectiveness of APG in common pattern modeling.
Furthermore, sharing parameter also contributes to the efficiency, due to fewer specific parameters generated (see Section \ref{sec:Efficiency Evaluation}). 

\textbf{The impact of the over parameterization.}
% In this part, we evaluate the effect of the proposed over parameterization.
Comparing the performance with or without over parameterization (i.e., v4 vs. v5), it shows that v5 performs better than v4 in all cases, which indicates adding more shared parameters can enrich the model capacity and lead to better performance.
% Besides, since there is no additional time and memory cost through over parameterization during inference (see Section \ref{sec:Over Parameterization}), such a nice property gives us more flexibility in the parameter space design.
% Furthermore, we extend over parameterization to a stack version which further improves the performance (see Appendix \ref{sec:An extension of the over parameterization}).

In addition, we also give a detailed analysis about the impact of the condition design in Appendix \ref{sec:Evaluation of the Different Aggregation Functions} and \ref{sec:Evaluation of Condition Design}, and the impact of the hyper-parameters in Appendix \ref{sec:Evaluation of the hyper-parameters}.
Furthermore, the influence to different frequency instances are also discussed in Appendix \ref{sec:Evaluation on the industrial application}

\begin{table}[]
% \begin{adjustbox}{width=\columnwidth,center}
\caption{The training time per epoch and memory cost for different versions of APG. 
$\Delta$ is the relative difference with respect to Base.
% The backbone architecture is 
% Note LP refers to the low-rank parameterization. 
% $|\Theta_i|$ and $|\Theta|$ refer to the specific and shared parameter size in a deep CTR model respectively.
% DFF refers to the decomposed feed-forwarding.
% We report AUC (\%) as the performance metric, second/epoch as the time cost, and megabytes as the memory cost, and the corresponding ratio compared with Base is also presented.
% Besides, in this part, to avoid the influence of over parameterization, all of the versions of APG do not adopt over parameterization.
}
\centering
\resizebox{\columnwidth}{!}{
\begin{tabular}{l|cc|cc|cc|cc|cc}
\toprule

\multirow{2}{*}{\textbf{Method}} & \multicolumn{2}{c|}{\textbf{[393,64,32,16]}}                              & \multicolumn{2}{c|}{\textbf{[393,128,64,32]}}                             & \multicolumn{2}{c|}{\textbf{[393,256,128,64]}}                           & \multicolumn{2}{c}{\textbf{[393,512,256,128]}}                            & \multicolumn{2}{c}{\textbf{[393,1024,512,256]}}       \\  \cmidrule{2-3}  \cmidrule{4-5}  \cmidrule{6-7}  \cmidrule{8-9} \cmidrule{10-11} 
{}                        & \textbf{Cost}  & \textbf{$\Delta$}   & \textbf{Cost}  & \textbf{$\Delta$}   & \textbf{Cost}  & \textbf{$\Delta$}  & \textbf{Cost}  & \textbf{$\Delta$} & \textbf{Cost}  & \textbf{$\Delta$}     \\ \midrule
\multicolumn{11}{c}{\textbf{Time / Epoch (s)}}                                                                                                                                                                                                                                                                                                        \\ \midrule
Base                     & {4.52}  & ---    & {5.27}  &---   & {5.89}   & ---   & {7.12}   &---   & {13.04}   & ---    \\\midrule
v1                      & {16.52} & {265.5\%}  & {50.31} & {854.6\%}  & {126.27} & {2043.8\%} & {420.33} & {5803.5\%} & {1462.78} & 11117.6\% \\ \midrule
v2             & {13.83} & {206.0\%}  & {39.53} & {682.8\%}  & {90.43}  & {1435.3\%} & {363.07} & {4999.3\%} & {1065.35} & 8069.9\%  \\\midrule
v3 & 6.31  &39.6\% &5.42 & 7.3\%&  5.71 & -3.1\% &6.31 & -11.4\% &11.23 &-13.9\% \\ \midrule
v4  &4.49  & -0.7\%  &4.78  & -5.3\%  &5.46  & -7.3\%  &5.89  & -17.3\%   &7.99   &-38.7\% \\ \midrule
\multicolumn{11}{c}{\textbf{Memory (M)}}                                                                                                                                                                                                                                                                                                              \\ \midrule
Base                     & {0.50}  & ---    & {0.89}  & ---   & {1.93}   & ---   & {4.61}   & ---    & {12.91}   & ---     \\ \midrule
v1                     & {11.21} & {2160.1\%} & {24.20} & {2613.0\%} & {56.32}  & {2818.1\%} & {144.84} & {3041.9\%} & {419.11}  & 3146.4\%  \\\midrule
v2                  & {0.74}  & {50.0\%}   & {0.91}  & {1.7\%}    & {1.21}   & {-37.3\%}  & {1.98}   & {-57.0\%}  & {3.22}    & -75.1\%   \\ \midrule
v4                 & {0.27}  & {-45.0\%}  & {0.29}  & {-68.0\%}  & {0.31}   & {-84.1\%}  & {0.35}   & {-92.4\%}  & {0.44}    & -96.6\%   \\ \bottomrule
\end{tabular}
}
\label{table:The efficiency results about different versions of APG.}
% \end{adjustbox}
\vspace{-1em}
\end{table}

% \vspace{-1em}
\subsection{Efficiency Evaluation}
\label{sec:Efficiency Evaluation}
% \vspace{-1em}
In this section, we evaluate the time and memory efficiency of our proposed method.
To this end, we train different versions (see Table \ref{table:The settings of different APG versions. }) of APG on the dataset IAAC and analysis the influence of each extension introduced in Section \ref{sec:Effective and Efficient Adaptive Parameter Generation Network}.
Since over parameterization does not introduce any cost, it is not considered here.
The decomposed feed-forwarding does not bring extra memory cost, it is not discussed in the memory usage.
For all versions, we set $K=4$ and the backbone (also the base) is WDL with 3 hidden layers.
We gradually increase the number of hidden units from [64,32,16] to [1024,512,256] with a fixed input\_shape=393 to evaluate the memory and time cost in different model scales.
% The results are presented in Table \ref{table:The efficiency results about different versions of APG.}
In Table \ref{table:The efficiency results about different versions of APG.}, we report the training time per epoch, memory usage\footnote{Since in this paper we mainly focus on the improvement of the hidden layers, we only count the memory cost of these hidden layers and the other parts (e.g., embedding layers) are not included here.} of each version.
% the relative difference with respect to base.

% Our approaches have several attractive properties.
% Based on our analysis in Section \ref{sec:Evaluation the performance with standard deep CTR models} and \ref{sec:Effectiveness Evaluation}, the proposed APG can significantly improve the performance.
% In addition to the effectiveness of APG, we

For the basic model (v1), although it achieves high performance (see Table \ref{table:The AUC results of the evaluation about different versions of APG. }), it is time expensive (265.5\% $\sim$ 11117.6\% relative to Base) and  memory costly (2160.1\% $\sim$ 3146.6\% relative to Base).
% Even worse,  the time cost of v1 is sharply increased with the increase of the backbone scales (from [64,32,16] to [1024,512,256]).
% although theoretically the time cost of v1 is  scales linearly with Base (i.e., $\mathcal{O}\textit{(NMD)$ in v1 vs. $\mathcal{O}\textit{(NM)$ in Base),
% It indicates that the computation bottleneck exacerbates the inefficiency problem, especially for a large scale backbone.
Such an inefficient model can hardly be accepted in web-scale applications.

% When introducing the low-rank parameterization (see v2 in Table \ref{table:The efficiency results about different versions of APG.}), the inefficiency problem in basic model is significantly addressed across all cases.
% Moreover, compared with Base, v1 can also reduce memory usage (e.g., -75.1\% in the large scale backbone).
% Note when we set $N,M$ to a small value, i.e., adopting a small scale backbone (e.g., [64,32,16] or [128,64,32]), the memory cost, theoretically in $\mathcal{O}\textit{{(NK+MK)D}}$, is more sensitive with $K$ and $D$, leading to a little increasement.
% For the time efficiency,the contribution the low-rank parameterization can be summarized as follows:
% (1) Although v2 still needs high time requirement due to the weight matrix $\bm{W}_i$ reconstruction compared with v1, the overall time cost of v2 is decreased.
% (2) The low-rank parameterization naturally contributes to the decomposed feed-forwarding which plays a key role in efficient learning. 

When introducing the low-rank parameterization (see v2 in Table \ref{table:The efficiency results about different versions of APG.}), compared to the basic model, the inefficiency problem in v2 is addressed across all cases.
Moreover, compared with Base, v1 can reduce memory usage (e.g., -75.1\% in the large scale model).
Note when we set $N,M$ to a small value (e.g., [64,32,16] or [128,64,32]), the memory cost, theoretically in $\mathcal{O}\textit{{((NK+MK)D})}$, is more sensitive with $K$ and $D$, leading to a little increasement.
For the time efficiency, the contribution of the low-rank parameterization can be summarized as follows:
(1) Although v2 still needs a high time requirement due to the weight matrix $\bm{W}_i$ reconstruction, compared with v1, the overall time cost of v2 is decreased.
(2) The low-rank parameterization naturally contributes to the decomposed feed-forwarding which plays a key role in efficient learning.

For version v3 which adopts decomposed feed-forwarding, it is free from the high computation of reconstructing the weight matrix $\bm{W}_i$ and achieves great improvement  by -13.9\% in time cost when considering a large scale model.
In addition, since v3 does not introduce any extra memory cost, it has the same memory usage as v2.
Note in the small model, the computation is far from the GPU bottleneck and time cost is insensitive with GFlops, leading to less improvement. 
% training time is also more sensitive with $K$ and $D$, leading to less improvement. 

When introducing the sharing parameters, v4 only needs to generate a small weight matrix, where time and memory cost (i.e., $\mathcal{O}\textit{(KKD)}$) in the generation process are free from the scales of a model.
Thus, the memory requirement of v4 is best among all cases, reducing memory by -45\% to -96.6\% relative to Base.
v4 also speeds up training time substantially, by -0.7\% to -38.7\% relative to Base.
It also supports the theoretical complexity analysis in Section \ref{sec:Model Complexity}.
 % where the time and memory cost is $\mathcal{O}\textit{(KKD+NK+MK)}$ in v4 while the corresponding cost in Base is $\mathcal{O}\textit{(NM)}$.
% In addition, 

Overall, such nice properties in terms of high performance, efficient memory usage, and low time requirement of our proposed model are welcomed for web-scale applications.

\subsection{Comparison with coarse-grained parameter allocating methods}
\label{sec:Comparison the performance with predefined-based methods}
% We also notice there are some works also focusing on developing custom parameters.
% In this section, we try to apply coarse-grained parameter allocating methods (including Star and MMoE) to fine-grained versions.
% % compare the proposed method with these coarse-grained parameter modeling methods.
% Since the specific parameters of Star and MMoE should be pre-defined manually, we allocate the instances (containing different items) with different parameters.
% To have a fair comparison, we also set APG with the group-wise condition and generate different parameters for the instance with different items.
% % we also apply these methods to CTR prediction tasks.
% % Besides, to have a fair comparison, the backbone of these methods is set the same, i.e., adopting WDL (which is a widely used backbone in practice) as the backbone.
% The AUC and Memory results are reported in Table \ref{table:The comparison with different coarse-grained parameter modeling methods.}.
% It can be found that compared with base, Star and MMoE can only achieve slight improvement but have huge memory costs.
% It indicates that such pre-defined and coarse-grained allocating methods are not suitable for fine-grained parameters modeling.

In this section, we compare the performance with coarse-grained parameter allocating methods (CGPMs), including Star and MMoE.
Specifically, we conduct two settings (see Table \ref{table:The comparison with different coarse-grained parameter modeling methods.}), i.e., coarse-grained (cg)  and fine-grained (fg), for each method.
For the former, methods adopt coarse-grained parameter modeling strategies, e.g., developing different parameters for different movie genders in MovieLens.
For the latter, methods adopt fine-grained parameter modeling strategies, e.g., developing different parameters for different movies in MovieLens.
The results are presented in Table \ref{table:The comparison with different coarse-grained parameter modeling methods.}.
% Some observations can be summarized as: 
% (1) Compared with base, although Star and MMoE can achieve some improvement, they need huge memory costs due to storing different parameters of different instances.
We can find, compared with Star and MMoE, APG shows superiority in efficiency and effectiveness.
The reasons are that (1) the specific parameters of APG are dynamically generated on-the-fly, without any need to store these specific parameters, and the architecture of APG is also designed efficiently to further save memory.
For Star and MMoE, they have to store these specific parameters due to the manually allocating strategies used in CGPMs.
Even worse, when comes to a fine-grained setting, more parameters are maintained.
(2) APG has better generalization than Star and MMoE, leading to better performance.
Actually, APG adopts parameter generation manner which provides a potential opportunity for generalization, and it has the ability to imply the inherent similarity between different specific parameters (see Section \ref{sec:Visualization} for additional experiments). 
Especially for the fine-grained setting where there may be a lack of enough instances to train the specific parameters,  the inherent connections among different specific parameters are helpful for parameter learning.

\subsection{Visualization}
\label{sec:Visualization}
\begin{wrapfigure}{r}{0.4\textwidth}
\centering
\includegraphics[width = .4\textwidth,scale=0.8]{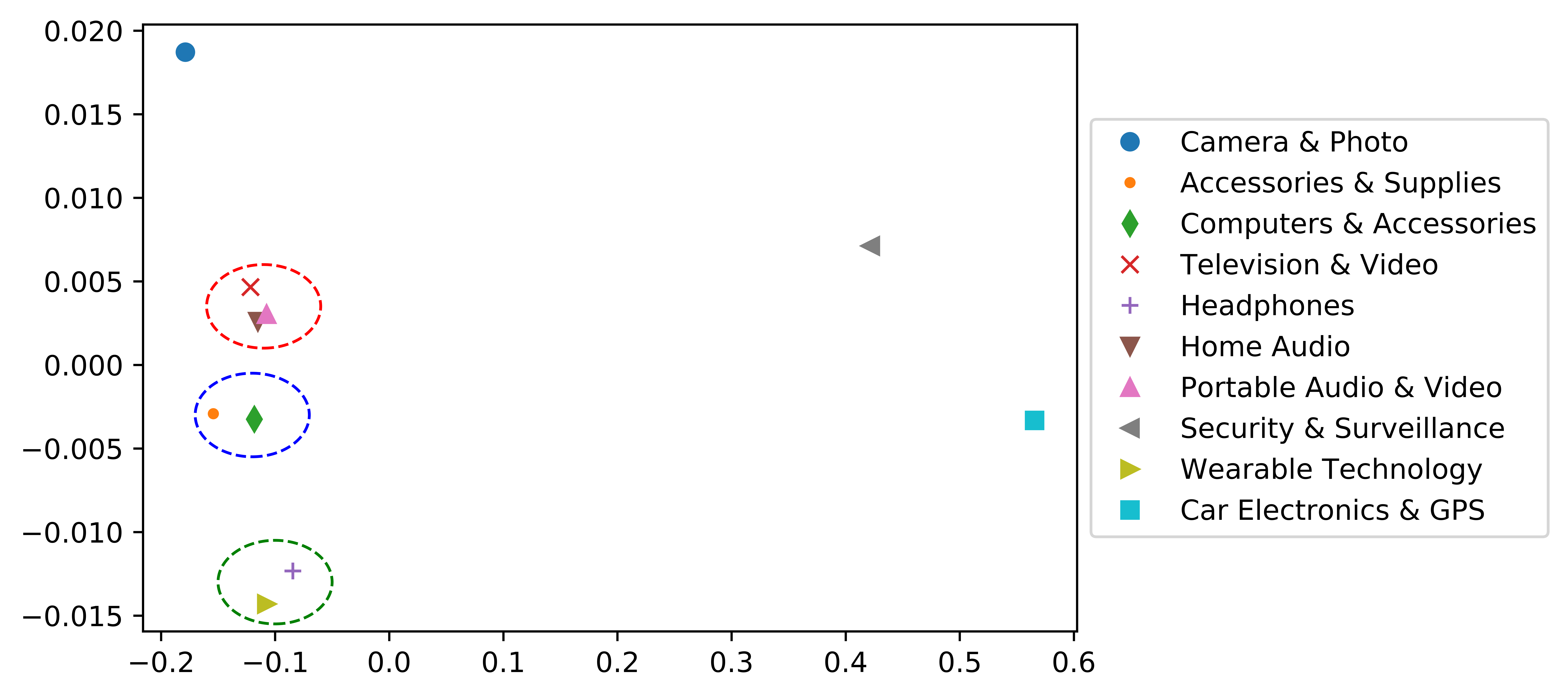}
\vspace{-1em}
\caption{The visualization of the generated specific parameters.
}
\vspace{-1em}
\label{figure:The visualization of the generated specific parameters}
\end{wrapfigure}
In this section, we visualize the generated specific parameters.
Specifically, we take the group-wise strategy on Amazon as an example.
The item category is set as the condition, and there is a total of 10 different categories.
Then we plot the generated specific parameters (i.e., $\bm{S}_i$) into a 2-D space by PCA \cite{wold1987principal}.
The visualization is presented in Figure \ref{figure:The visualization of the generated specific parameters}.
Each point refers to the specific parameters generated for one specific category.
Interestingly, the observed groupings (i.e, in the same dashed circle) correspond to similar categories. 
This shows that the learned specific parameters by APG are meaningful and can capture the relations among different specific parameters implicitly.

\begin{table}[t]\small
\caption{The comparison with coarse-grained parameter allocating methods. 
Mem refers to the memory cost (M) of the hidden layers.
}
\centering
\resizebox{\columnwidth}{!}{
\begin{tabular}{l|l|l|l|l|l|l|l|l|l}
\toprule
\multirow{2}{*}{} & \multicolumn{3}{c|}{\textbf{MovieLens}}                        & \multicolumn{3}{c|}{\textbf{Amazon}}                           & \multicolumn{3}{c}{\textbf{IAAC}}                             \\ \cmidrule{2-10} 
                  &Setting &{AUC}   & Mem  & Setting& {AUC}   &Mem & Setting & {AUC}   & Mem \\  \midrule
Base              & ---     &{79.21} & 1.29    &--      &{69.15} & 1.51     & --- & {65.17} & 1.93                     \\  \midrule
Base+MMoE$_{fg}$  & movie   & {79.31} & 123.42 &item    &{69.17} & 315.93   &item & {65.22} & 332.97                    \\
Base+MMoE$_{cg}$  & movie gender  & 79.28 &  32.91   &item category&69.14 & 5.78       &item brand             & 65.20 & 70.53                   \\ \midrule

Base+Star$_{fg}$  & movie   &{79.28} & 3784.23    &item              & {69.20} & 11034.02        &item          & {65.36} & 13753.71                  \\  
Base+Star$_{cg}$  & movie gender  &  79.25       & 311.56  &item category                & 69.20 & 42.84      &item brand          & 65.28& 2866.88                  \\  \midrule
Base+APG$_{fg}$   & movie   & \textbf{79.58} & 0.24 &item                    & \textbf{69.35} & 0.26  &item                  & \textbf{65.76} & 0.38                     \\ 
Base+APG$_{cg}$   & movie gender  &  79.46       & 0.24    &item category & 69.30 &          0.26        &item brand     & 65.59&       0.38              \\   \bottomrule

\end{tabular}
}
\label{table:The comparison with different coarse-grained parameter modeling methods.}
\end{table}

% \begin{table}
% \caption{The industrial results}
% \begin{tabular}{l|c|c|c}
% \toprule
%     & \textbf{AUC}  & \textbf{CTR} & \textbf{RPM} \\ \midrule
% Gains & +0.2\% & +3\% &  +1\%    \\ 
% \bottomrule
% \end{tabular}
% \label{table:The industrial results}
% \end{table}

\section{Conclusion}
\label{sec:Conclusion}
In this paper, we propose an efficient, effective, and universal module to adaptively generate parameters for different instances.
In this way, the model can carefully characterize the patterns for different instances by adopting different parameters.
% At the same time, since the adaptive parameters are generated on-the-fly, there is no need to store these generated parameters, which significantly saves memory.
Experimental results show that with the help of APG, all of the existing deep CTR models can make great improvements, which also encourages a wide application for APG.
Furthermore, the effectiveness and efficiency of APG are also detailed analyzed.
Currently, APG requires users to set some hyper-parameters, e.g., condition strategies, $K$, $P$, and etc.
In the future, we will attempt to automatically implement APG with different settings for different situations.

\appendix
% \section*{Appendix}

\section{Examples for different condition designs}

\subsection{Group-wise}
\label{sec:group-wise example}
\begin{exmp} [\textbf{Thousand users with Thousand Models (a.k.a. Personalized Parameters)}]
In a traditional recommendation model, although some personalized signals (e.g., user-item interactions) are introduced, the shared parameters cannot significantly characterize the personality of each user especially for long-tailed users.
In our method, with the help of the group-wise strategy, we can simply give the prior knowledge about users to the model parameters to capture personalized patterns and achieve "Thousand users with Thousand Models".
Specifically, given the embedding of a user $m$ as $\bm{u^m}_i \in \mathbb{R}^d$ and an involved instance $i$ of this user, we can directly set $\bm{z}_i=\bm{u^m}_i$.
In other words, by the prior knowledge $\bm{u^m}_i$, we explicitly group the instances by users and allow different users to enjoy different parameters.
\end{exmp}

\subsection{Mix-wise}
\label{sec:mix-wise example}
\begin{exmp} [\textbf{Real-time Personalized Parameters}]
\label{exmp:Real-time Personalized Parameters}
In practice, user interests may be dynamically changed.
% Simply applying group-wise strategy can hardly capture such changes in parameters space.
For a mix-wise strategy, we can add prior knowledge about users' latest behaviors to allow the parameters sensitive to real-time interests.
Specifically, given an instance $i$ associated with a user $m$, we denote $\bm{z^{0,m}}_i \in \mathbb{R}^d$ as the user $m$ embedding and $\bm{z^{1,m}}_i \in \mathbb{R}^d$ as the latest behaviors embedding of user $m$ \footnote{The latest behaviors embedding can be obtained by averaging the embedding of the latest clicked items of this user}.
Then these two conditions $\bm{z^{0,m}}_i,\bm{z^{1,m}}_i$ are both considered to generate parameters.
In this way, the generated parameters are specific for different users and can be adjusted in real-time by different user interests.
\end{exmp}

\begin{exmp} [\textbf{Thousand instances with Thousand Models (a.k.a. Instance-aware Parameters}]
We can consider more conditions to allow the model parameters sensitive to the instance level.
To achieve this goal, the conditions are required to identify each instance.
Specifically, assuming in a recommendation application, for each instance, we can take the embedding of the associated user, item, and context as $\bm{z}^j_i$.
\end{exmp}

\section{The detailed experimental setting}
\label{sec:The detailed experimental setting}

\subsection{Datasets}
Four real-world datasets are used:
% \begin{itemize}
% \item  \textbf{Amazon.} \footnote{https://www.amazon.com/} is collected from the Electronics category on Amazon.
% There are total 1,292,954 instances, 1,157,633 users.
% \item \textbf{MovieLens.} \footnote{https://grouplens.org/datasets/movielens/} is a reviews dataset and is collected from the MovieLens web site.
% There are total 1,000,209 instances, 6,040 users.
% \item \textbf{IJCAI2018 Advertising Algorithm Competition (IAAC).}\footnote{https://tianchi.aliyun.com/competition/entrance/231647/introduction} is a dataset collected from a sponsored search in E-commerce.
% Each record refers to whether a user purchases the displayed item after clicking this item.
% There is a total of 478,138 records, 197,694 users, and 10,075 items.
% \item  \textbf{Alibaba.} is an industrial dataset which is obtained from Taobao used for industrial evaluation (see Section \ref{sec:Results on industrial application}). 
% Each instance refers to a user who searches a query in this platform, and the platform returns an item to this user.
% The label is defined as whether the user clicks this item.
% There are a total of 4 billion instances, 100 million users.
% \end{itemize}
(1)\textbf{Amazon} \footnote{https://www.amazon.com/} is collected from the electronics category on Amazon.
There are total 1,292,954 instances, 1,157,633 users.
(2) \textbf{MovieLens.} \footnote{https://grouplens.org/datasets/movielens/} is a review dataset and is collected from the MovieLens web site.
There are total 1,000,209 instances, 6,040 users.
(3) \textbf{IJCAI2018 Advertising Algorithm Competition (IAAC)}\footnote{https://tianchi.aliyun.com/competition/entrance/231647/introduction} is a dataset collected from a sponsored search in E-commerce.
Each record refers to whether a user purchases the displayed item after clicking this item.
There is a total of 478,138 records, 197,694 users, and 10,075 items.
(3) \textbf{Industrial dataset (IndusData)} is used for industrial evaluation (see Appendix \ref{sec:Evaluation on the industrial application}). 
Each instance refers to a user who searches a query in this platform, and the platform returns an item to this user.
The label is defined as whether the user clicks this item.
There are a total of 4 billion instances, and 100 million users.
% \end{itemize}
The statistics of the data sets are summarized in Table \ref{table:The statistic of datasets}.

\begin{table}[h]
\center
\caption{The statistic of datasets.}
% \vspace{-1em}
\begin{tabular}{l|c|c|c}
\toprule
        & \textbf{\#Data}    & \textbf{\#User ID}   & \textbf{\#Item ID}   \\\midrule
Amazon &1,292,954 &1,157,633& 9,560\\ \midrule
MovieLen & 1,000,209 &  6,040 & 3,706\\ \midrule
IAAC & 478,138 &  197,694 & 10,075 \\ \midrule
IndusData & 4 billion & 100 million & 80 million  \\ 
\bottomrule
\end{tabular}
% \vspace{-1em}
\label{table:The statistic of datasets}
\end{table}

\subsection{Baselines.}
\label{sec:Baselines.}
Here, we compare our method with two kinds of methods

\noindent \textbf{Existing CTR prediction methods:} 
To show the effectiveness of the proposed APG, we apply it to various existing deep CTR models
(1) WDL\cite{cheng2016wide} adopts wide and deep parts to memorize and  generalize patterns of instances.
(2) PNN\cite{qu2016product} explicitly introduces product operation to explore the interactions of categorical data in multiple fields.
(3) FIBINET\cite{huang2019fibinet} designs a squeeze-excitation network to dynamically learn the feature importance and use a bilinear-interaction layer to learn the interactions among features.
(4) DIFM\cite{lu2020dual} brings the idea of the transformer and learns vector-wise and bit-wise interactions among features.
(5) DeepFM\cite{guo2017deepfm} takes the linear part of WDL with an FM network to better represent low-order features.
(6) DCN\cite{wang2017deep} learns low-order and high-order features simultaneously and needs low computation cost.
(7) AutoInt\cite{song2019autoint} learns the feature interactions automatically vim self-attention neural networks.

\noindent \textbf{Coarse-grained parameter allocating methods:}
We also try to compare the proposed APG with the coarse-grained parameter allocating methods:
(1) Multi-task learning: MMoE \cite{ma2018modeling}  keeps multiple parameters by adopting multiple network branches for different tasks;
(2) Multi-domain learning: Star \cite{sheng2021one} allocates multiple parameters for different scenarios.

% \begin{itemize}
%  \item Multi-task learning: MMoE \cite{ma2018modeling}  keeps multiple parameters by adopting multiple network branches for different tasks. 
%  \item Multi-domain learning: Star \cite{sheng2021one} develops multiple parameters for different scenarios.
% \end{itemize}

% In order to have a fair comparison, 
% \subsection{Training Details.} 
% All methods have the same EDRM architecture.
\subsection{Training Details}
\label{sec:Training Details}
The embedding dimension is set 32 for all methods.
The number and units of hidden layers are set $\{256,128,64\}$ for all methods by default.
Other hyper-parameters of different methods are set by the suggestion from original papers.
The backbone of the deep CTR models is set as WDL by default.
For APG, we set the self-wise strategy as the default condition strategy.
$\mathcal{G}$ is implemented as an MLP with a single layer by default.  
We set the hyper-parameters $K \in \{2,4,6,8\}$ and $P \in \{32,64,128,256,512\}$  and perform grid search over $K$ and $P$.
We include the results for different values of $K$ and $P$ in Appendix \ref{sec:Evaluation of the hyper-parameters}.
We use the Adam optimizer  with a learning rate of 0.005 for all methods.
The batch size is 1024 for all datasets.
Each dataset is randomly split into 80\% train, 10\% validation, and 10\% test sets. 
For the public datasets, methods are trained on a single V100S GPU. 
For the industrial dataset, methods are trained in an internal cluster equipped with V100S GPU and SkyLake CPU. 
We run all experiments multiple times with different random seeds and report the average results.

\section{Evaluation of Condition Design}
\label{sec:Evaluation of Condition Design}
In this section, we analyze the influence of the condition design.
Specifically, we evaluate different condition strategies on CTR prediction tasks and report the AUC results. 
We take WDL as the backbone of APG.
For the mix-wise strategy, the input aggregation with attention function is used and the effect of the aggregation method of the mix-wise strategy is detailedly analyzed in Appendix \ref{sec:Evaluation of the Different Aggregation Functions}. 
Besides we also provide the results of WDL (defined as Base) for comparison.

\begin{table}[t] \small
\center
\caption{The AUC (\%) results of APG with different kinds of condition strategy, including Group-wise, Mix-wise, and Self-wise. 
Note Base refers to the results of the method without APG.
U, I, and C refers to the embedding of users, items,  and contexts respectively.}
\begin{tabular}{l|l|c|c|c}
\toprule
  \textbf{Strategy}                            & \textbf{$\bm{z}_i$}       & \textbf{MovieLens} & \textbf{Amazon} & \textbf{IAAC}                                       \\  \midrule
Base                         &                 & 79.21  & 69.15  & 65.17                                      \\ \midrule
                             & U               & 79.61  & 69.28  &65.80 \\ 
\multirow{-2}{*}{Group-wise} & I               & 79.58  & 69.35  &  65.76 \\ \midrule
%                              & U,B (i.e. Example 2)          &        &        &    \\ 
% \multirow{-2}{*}{Mix-wise}   & U,I,C (i.e. Example 3)& 79.45  & 69.31  & \textbf{65.94}   
Mix-wise   & U,I,C & 79.45  & 69.31  & 65.90                                     \\ \midrule
Self-wise                    & $\bm{x}_i$            & \textbf{79.73}  & \textbf{69.43}  &  \textbf{65.94} \\ 
\bottomrule
\end{tabular}
\label{table:Evaluation of the Condition Design}
\end{table}

% \subsubsection*{\normalfont{\textbf{The effect of different condition strategies.}}}
\textbf{The effect of different condition strategies.}
We first analyze the effect of different condition strategies, including the group-wise, the mix-wise, and the self-wise strategies.
% The results are reported in Table \ref{table:The Effect of different parameter strategy}.
From Table \ref{table:Evaluation of the Condition Design}, we can find that 
(1) Compared with Base, all of these condition strategies can obtain better performance.
It indicates the effectiveness of APG which can dynamically generate the parameters for better pattern learning;
(2) The self-wise strategy achieves the best performance among other strategies in all cases.
One of the possible reasons is that the prior knowledge of the self-wise strategy is directly from the hidden layers' input which is a more immediate signal for the current layer and may lead to better parameter generation.
% (3) Compared with other strategies, the group-wise strategy achieves high performance on MovieLens and Amazon and the mix-wise strategy obtains high AUC on IAAC.
% These indicates that 

% \begin{table}[t]
% \caption{The evaluation results of the over parameterization.
% % Note all version has the same shape (i.e., $\mathbb{R}^{N \times M}$) of the final adweight matrix.
% }
% \begin{tabular}{l|l|c|c|c}
% \toprule
% AUC(\%)     &\textbf{Version}& \textbf{MovieLens} & \textbf{Amazon} & \textbf{IAAC}  \\ \midrule
% Base       &$\bm{W}$ & 79.21     & 69.15  & 65.17 \\  \midrule
% % v1&$\bm{W}_i$        & 79.51    & 69.33  & 65.52 \\  \midrule
% % \multirow{2}{*}{
% v3&$\bm{U}\bm{S}_i\bm{V}$      & 79.56  &69.34  &65.67 \\  \midrule
% % v2&$\bm{U}\bm{S}_i\bm{V}$    , $\bm{S}_i \in \mathbb{R}^{K \times K}$   & 79.58 &\textbf{69.40} &66.04\\  \midrule
% % v3&$\bm{U}_i\bm{S}_i\bm{V}_i$  , $\bm{S}_i \in \mathbb{R}^{N \times M}$   & &&   \\  \midrule
% v6&$\bm{U}^l\bm{U}^r\bm{S}_i\bm{V}^l\bm{V}^r$     & 79.60     & 69.37  & 65.94 \\  \midrule
% % \multirow{2}{*}
% % v7&{${(\prod_{j}^{J}\bm{U}^j)\bm{S}_i(\prod_{j}^{J}\bm{V}^j)}$}, $J=3$ & 79.59     & 69.39  & \textbf{66.13} \\ \midrule
% % v8&{${(\prod_{j}^{J}\bm{U}^j)\bm{S}_i(\prod_{j}^{J}\bm{V}^j)}$},  $J=4$  & \textbf{79.61}     & 69.39  & 65.96 \\ \bottomrule
% \end{tabular}
% \label{table:Evaluation of the Re-Parameterization}
% \end{table}

% \subsubsection*{\normalfont{\textbf{The effect of different prior knowledge in the same strategy.}}}
\textbf{The effect of different prior knowledge in the same strategy.}
We also evaluate the effect of different prior knowledge in the same strategy.
Specifically, considering the group-wise strategy, we compare the performance between the prior knowledge about the user embedding and the item embedding. 
% For the mix-wise strategy, we evaluate the performance of Example 2 and Example 3 mentioned in Section \ref{sec:Mix-wise}.
The results are reported in Table \ref{table:Evaluation of the Condition Design}.
We can conclude that 
(1) Different prior knowledge in the same strategy can get competitive performance;
(2) Designing a proper prior knowledge may achieve better performance.
For example, for group-wise strategy, taking the item embedding as the prior knowledge obtains higher AUC on Amazon.
While the prior knowledge about the user embedding performs better on IAAC and MovieLens.

% \section{Additional Results of the evaluation of the hyper-parameters}
% \label{sec:Additional Results of the evaluation of the hyper-parameters}
% Here, we present additional results about the performance with different $P$ and $K$ on IAAC and MovieLens in Figure \ref{figure:Evaluation of the hyper-parameters on IAAC and MovieLens}.
% Detailed analysis can be found in Section \ref{sec:Evaluation of the hyper-parameters}.

\section{Evaluation of the Different Aggregation Functions}
\label{sec:Evaluation of the Different Aggregation Functions}
In this section, we conduct experiments to evaluate the performance when using different aggregation functions for  the mix-wise strategy.
All of the cases use the embedding of users, items, and the context as the condition.
The results are reported in Table \ref{table:The Results of Using Different Aggratetion Function for the Mix-wise Strategy}.
We can observe that 
(1) Compared with other functions, the Attention function achieves the best performance in most cases due to the high representation power of the attention function;
(2) Compared with Output Aggregation, Input Aggregation performers better.
One of the possible reasons is that the relations among different conditions can be implicitly modeled through $\mathcal{G}$ while output aggregation simply summarizes different specific parameters.

\begin{table}[h]\small
\center
\caption{The results of using different aggratetion function for the mix-wise strategy.}
\begin{tabular}{l|l|c|c|c}
\toprule
AUC (\%)                                    & \textbf{Function} & \textbf{MovieLens} & \textbf{Amazon} & \textbf{IAAC}  \\ \midrule
Base                                &                      & 79.21     & 69.15  & 65.17 \\ \midrule
\multirow{3}{*}{Input Aggregation}  & Mean                 & 79.33     & 69.28  & 65.44 \\ 
                                    & Concat               & 79.38     & \textbf{69.36}  & 65.73 \\ 
                                    & Attention            & \textbf{79.45}     & 69.31  & \textbf{65.90} \\ \midrule
\multirow{3}{*}{Output Aggregation} & Mean                 & 79.30     & 69.34  & 65.44 \\  
                                    & Concat               & 79.33     & \textbf{69.36}  & 65.57 \\  
                                    & Attention            & 79.41     & 69.25  & 65.70 \\ \bottomrule
\end{tabular}
\label{table:The Results of Using Different Aggratetion Function for the Mix-wise Strategy}
\end{table}

\section{Evaluation of Hyper-parameters}
\label{sec:Evaluation of the hyper-parameters}
% trim=left bottom right top, clip
% trim=0 0 40 30, clip,
\begin{figure}[t]
% % \vspace{-1em}
  % \centering
% \begin{wrapfigure}{r}{0.7\textwidth}
 \centering
     \subfigure[AUC vs. P on MovieLens]{
    \includegraphics[width= .3\textwidth]{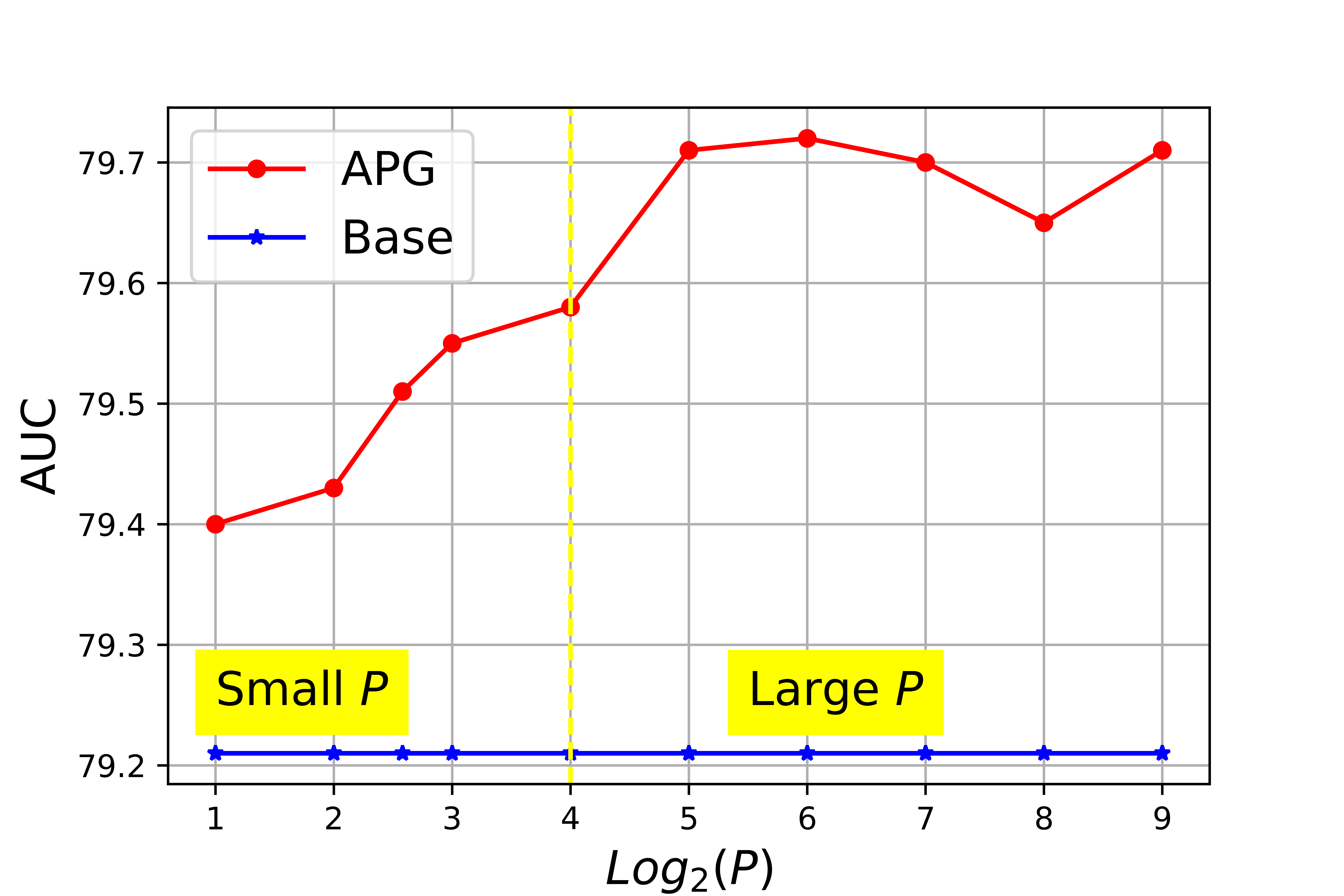}
    \label{fig:subfigure2}
  }
      \subfigure[AUC vs. K on MovieLens]{
    \includegraphics[ width= .3\textwidth]{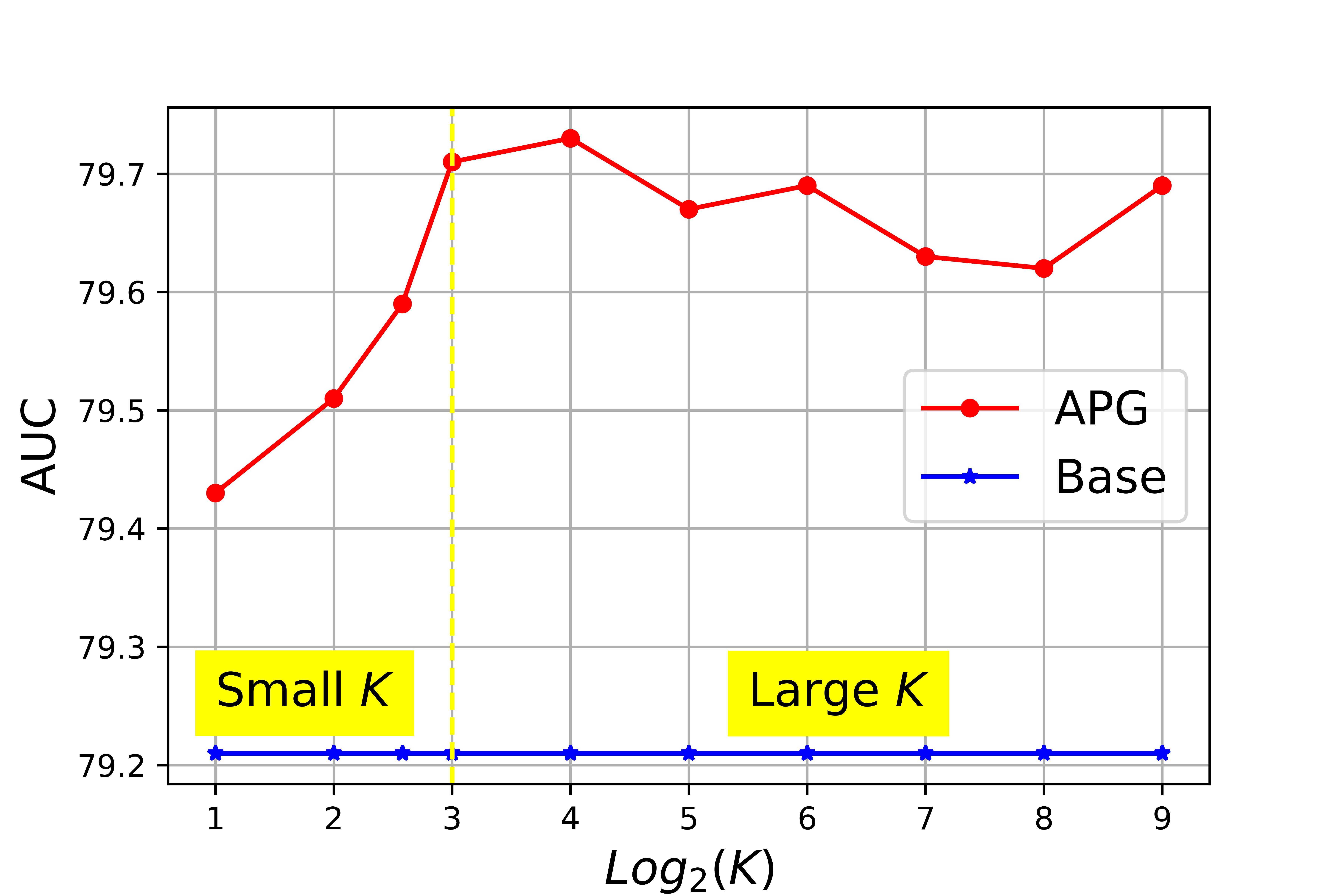}
    \label{fig:subfigure2}
  }
       \subfigure[AUC vs. \#layer on MovieLens]{
    \includegraphics[width= .3\textwidth]{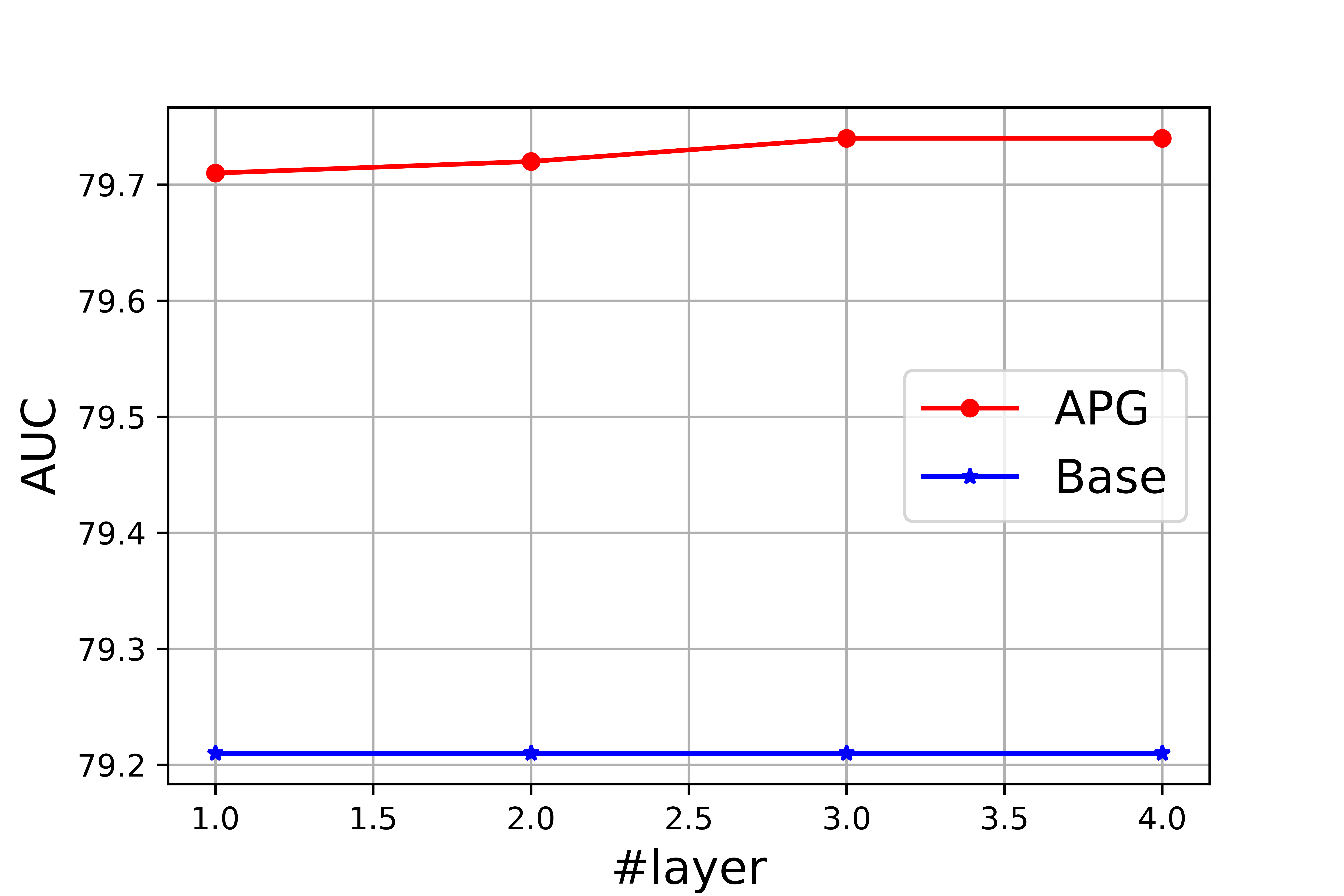}
    \label{fig:subfigure2}
  }

    \subfigure[AUC vs. P on Amazon]{
    \includegraphics[width= .3\textwidth]{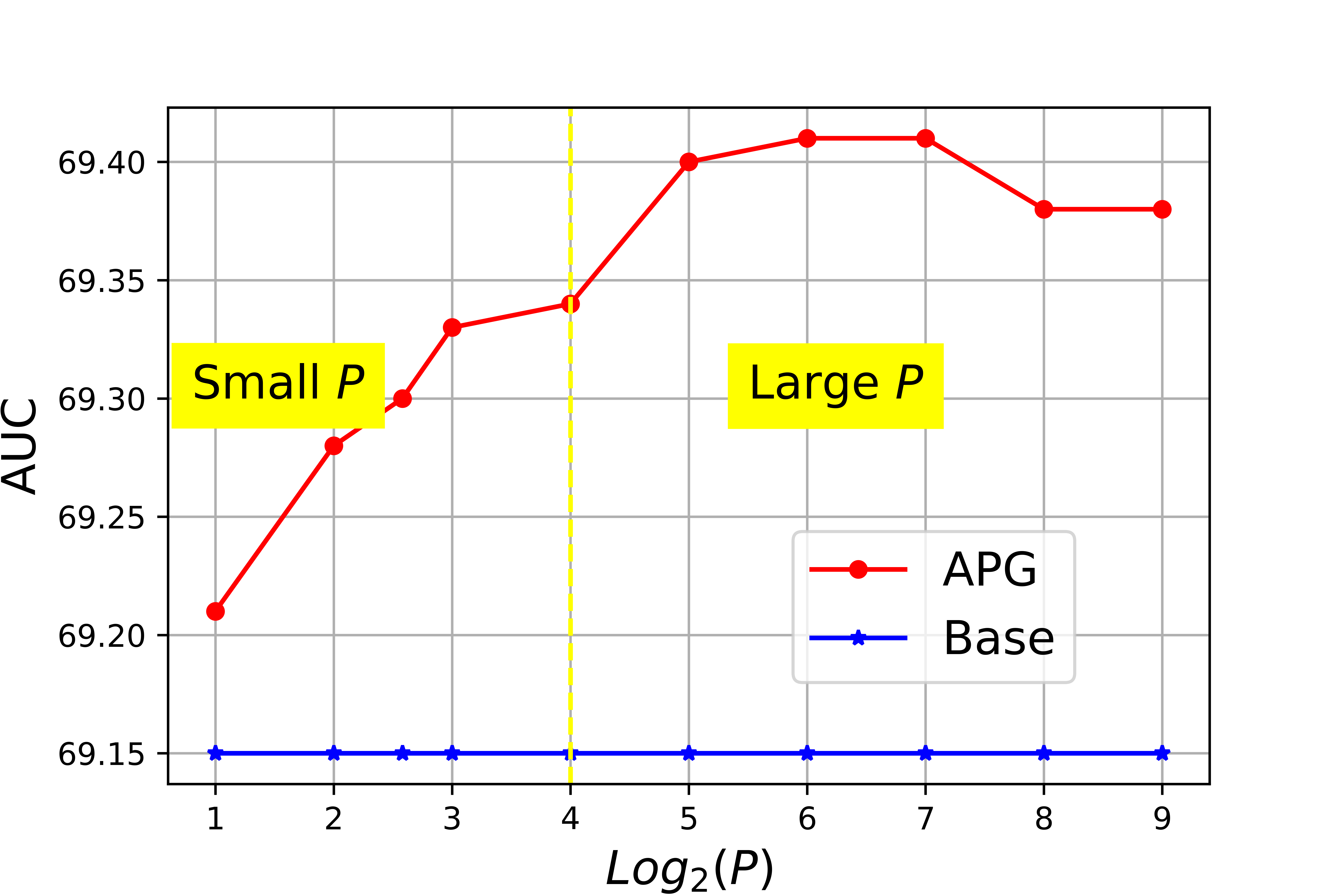}
    \label{fig:subfigure2}
  }
      \subfigure[AUC vs. K on Amazon]{
    \includegraphics[width= .3\textwidth]{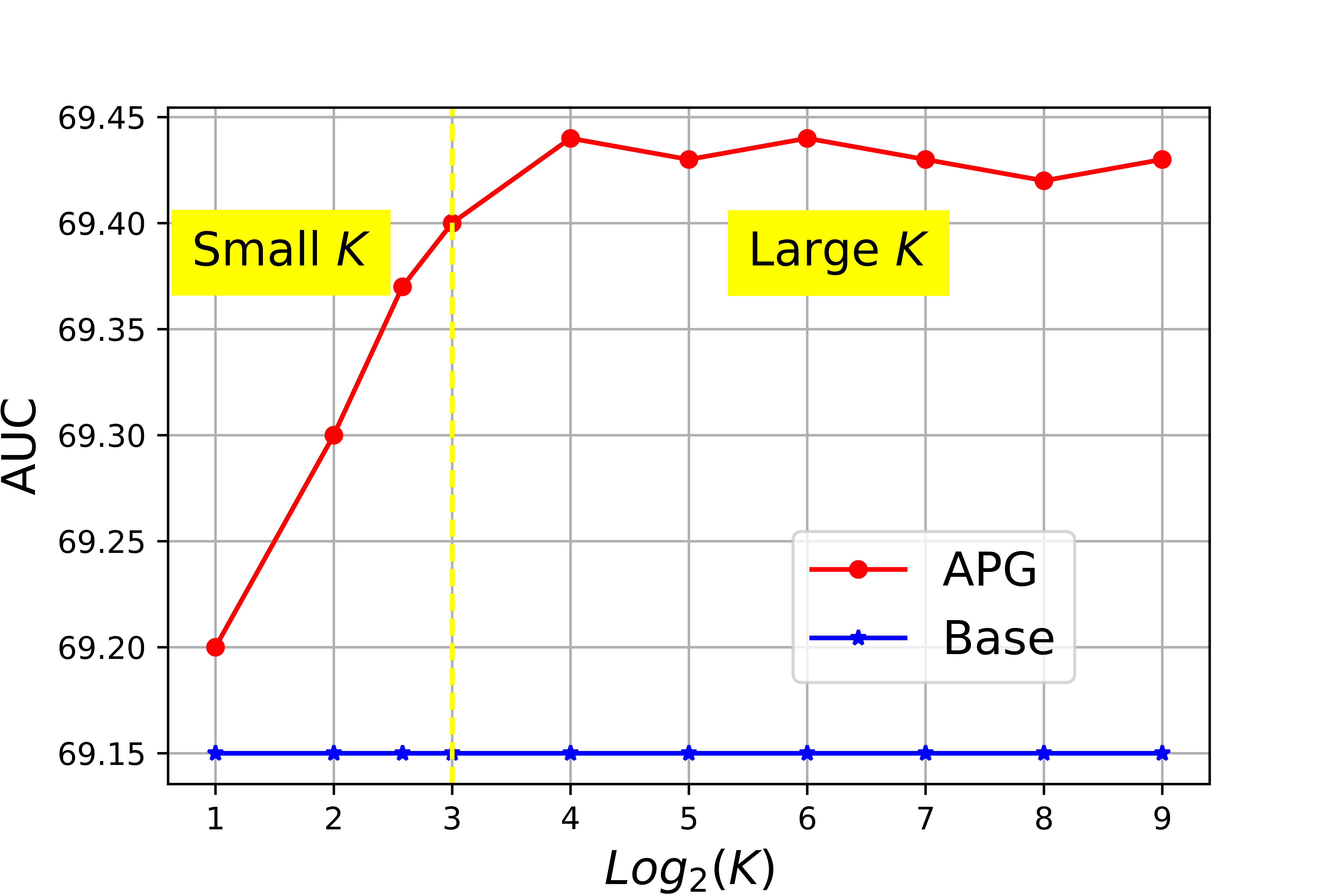}
    \label{fig:subfigure2}
  }
      \subfigure[AUC vs. \#layer on Amazon]{
    \includegraphics[width= .3\textwidth]{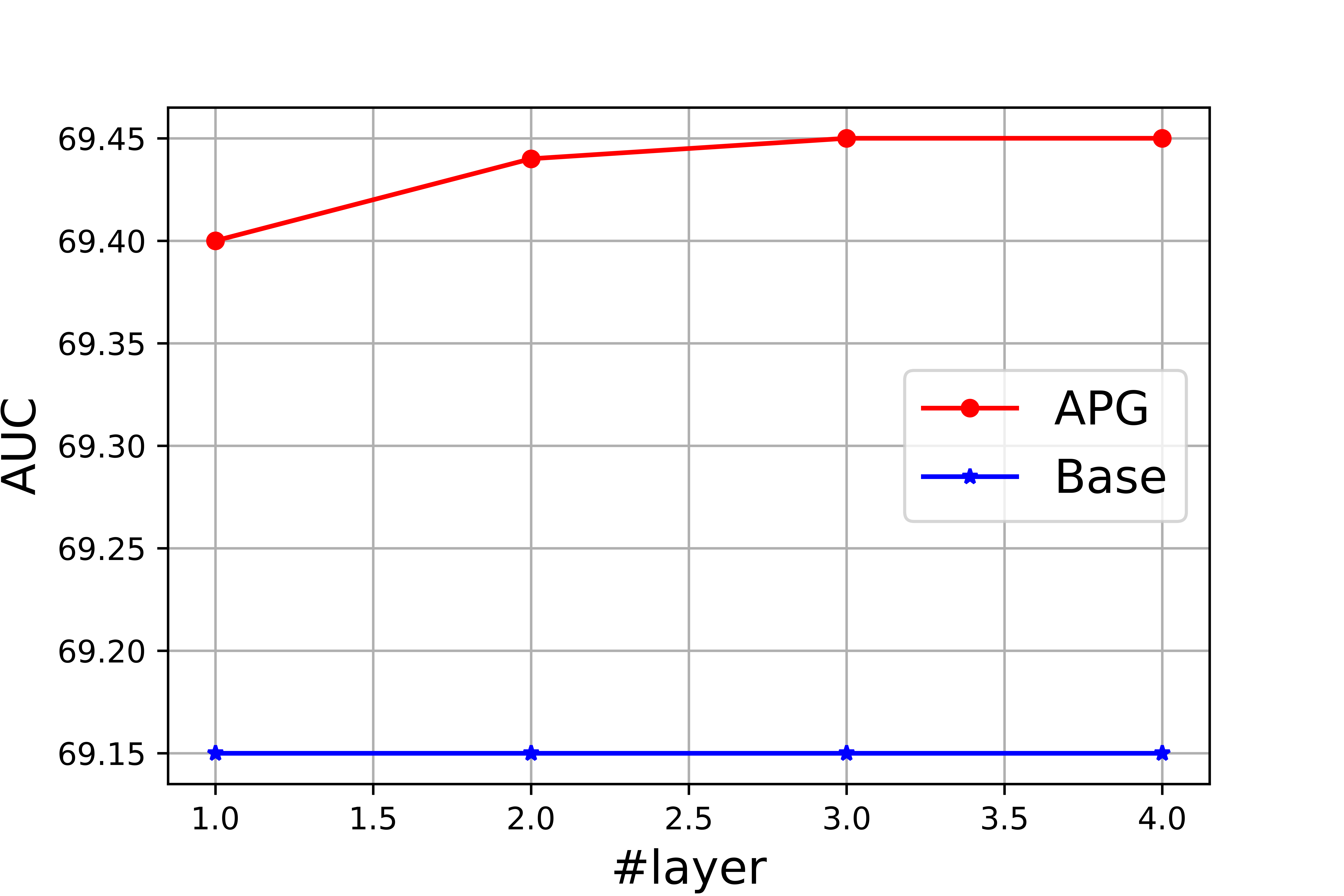}
    \label{fig:subfigure2}
  }

    \subfigure[AUC vs. P on IAAC]{
    \includegraphics[ width= .3\textwidth]{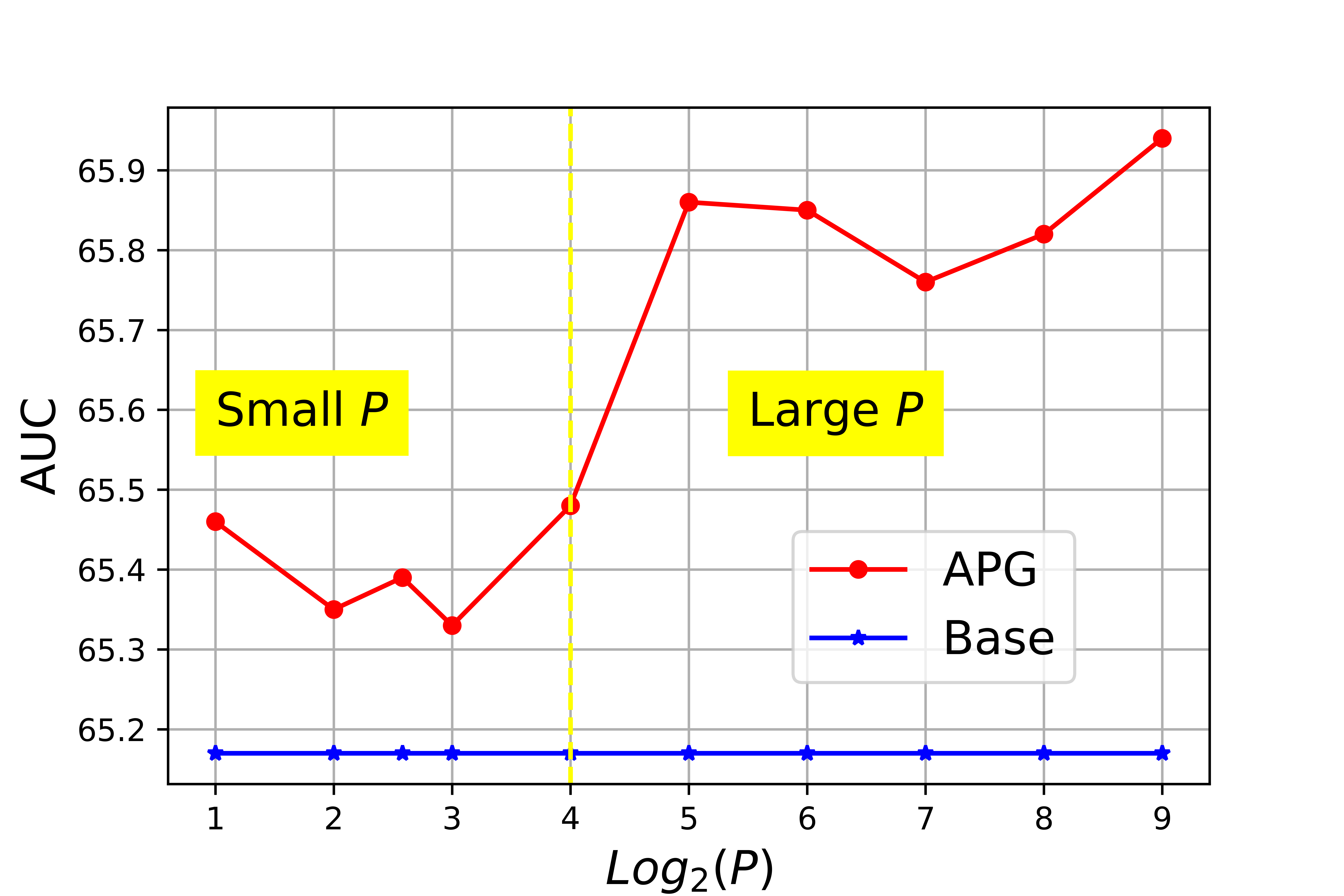}
    \label{fig:subfigure2}
  }
      \subfigure[AUC vs. K on IAAC]{
    \includegraphics[width= .3\textwidth]{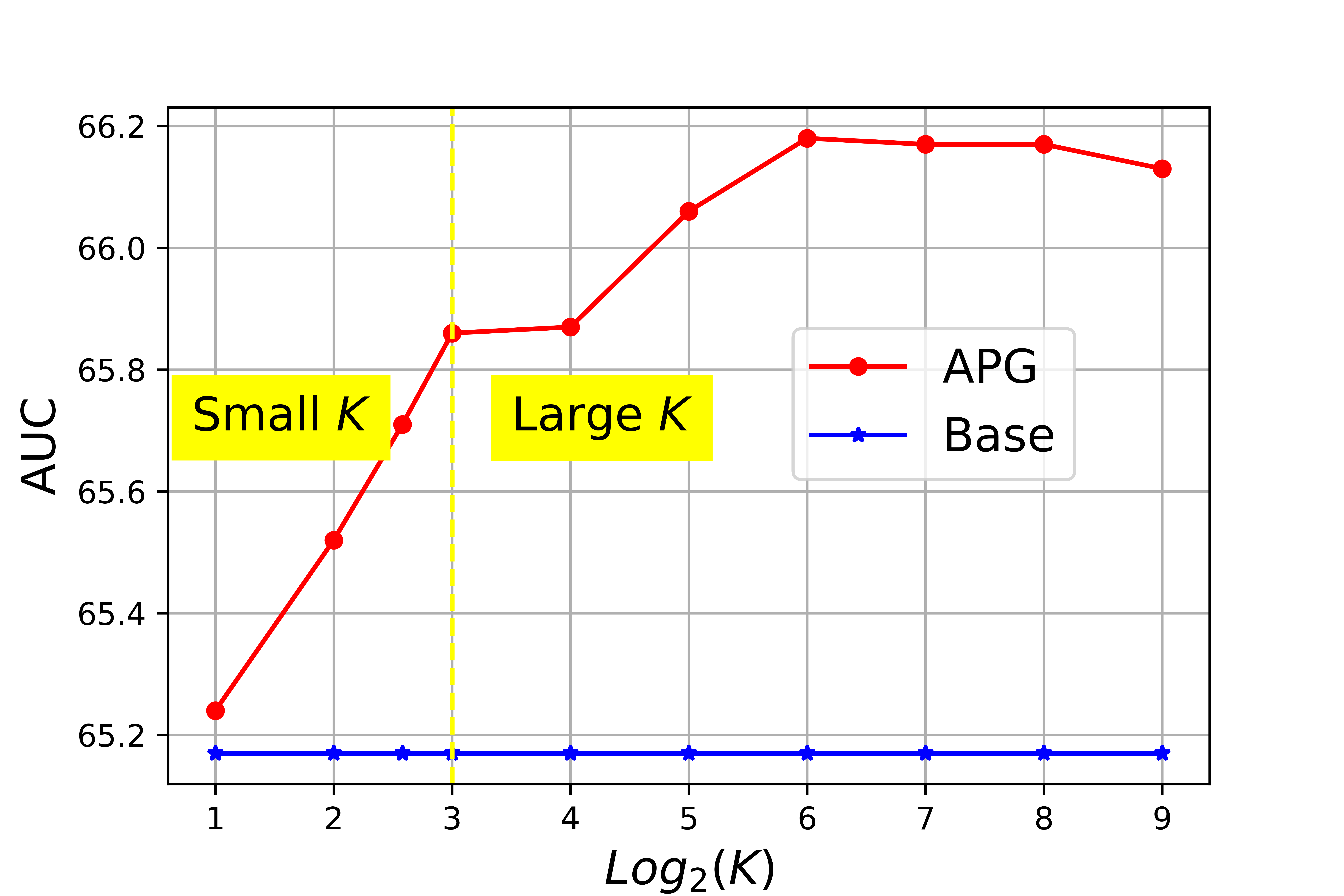}
    \label{fig:subfigure2}
  }
      \subfigure[AUC vs. \#layers on IAAC]{
    \includegraphics[width= .3\textwidth]{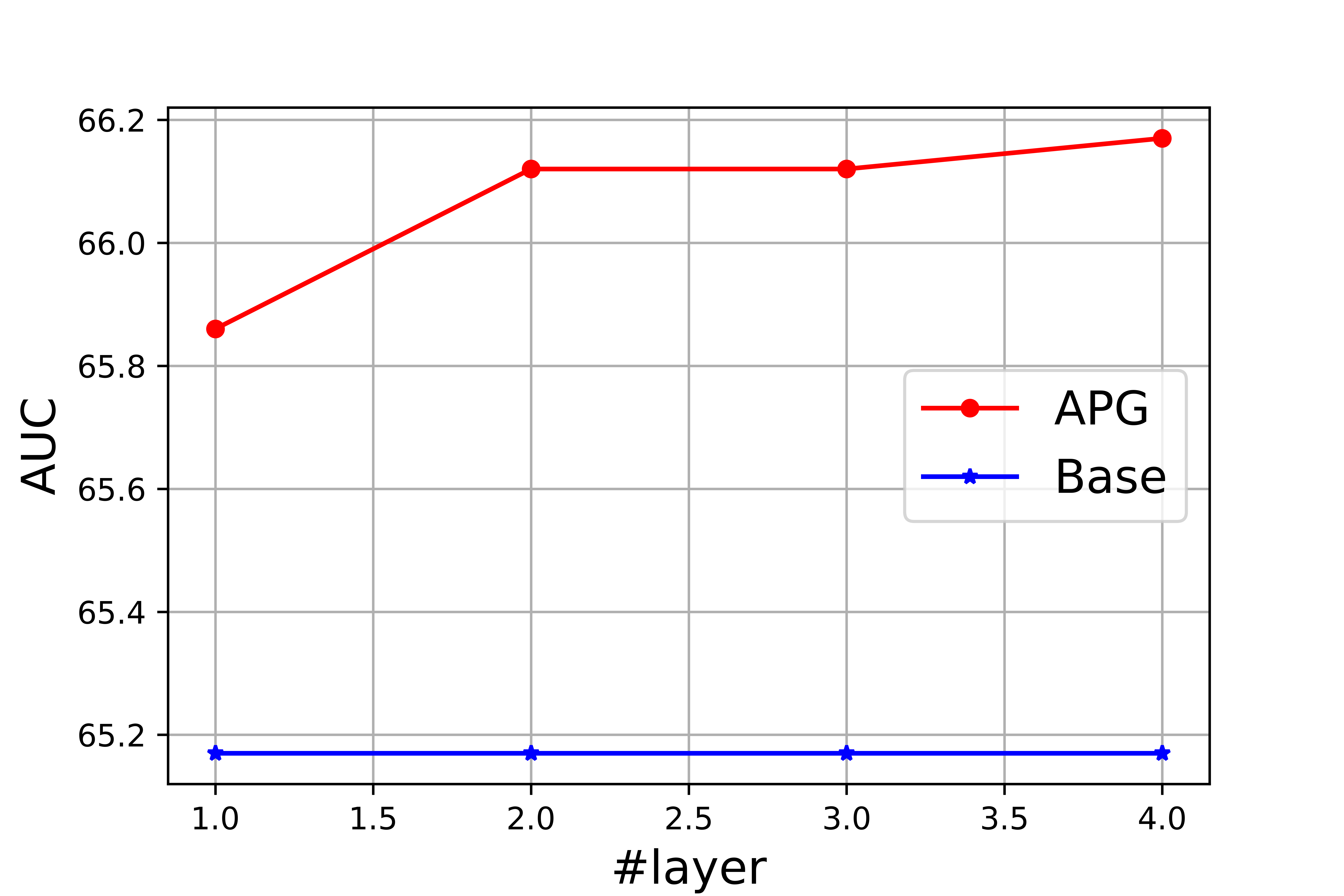}
    \label{fig:subfigure2}
  }
\caption{Evaluation of the hyper-parameters.}
% \vspace{-1em}
\label{figure:Evaluation of the hyper-parameters}
% \end{wrapfigure}
% \vspace{-1.5em}
% \caption{Evaluation of the hyper-parameters on Amazon.}
% % \vspace{-1em}
% \label{figure:Evaluation of the hyper-parameters}
\end{figure}
In this section, we conduct experiments to analyze the effect of the hyper-parameters including $P$, $K$, and the number of MLP layers in APG.
% Due to limited space, the results on Amazon are reported here and more results on other datasets can be found in Appendix \ref{sec:Additional Results of the evaluation of the hyper-parameters}.

\subsection{The effect of different $P$}
\label{sec:The effect of different P}
The hyper-parameter $P$ is introduced to add more shared parameters.
Thus $P$ is required to be much larger than $K$ (i.e., $P \gg K$).
Here we keep $K=8$ for all cases, and set $P$ to $\{32,64,128,256,512\}$ respectively and evaluate the performance of APG. 
The results are plotted in the right part (i.e., Large $P$) in Figure \ref{figure:Evaluation of the hyper-parameters} (a)(d)(g).
Note we also provide the results of the Base
 % and APG w/o over parameterization (denoted as APG ($\bm{US}_i\bm{V}$)) 
for comparison.
We can find that setting different large values of $P$ can give a similar better performance compared with Base.
It indicates that when adding more parameters by setting a large $P$, the model can be stably improved.

% \subsubsection*{\normalfont{\textbf{Does a small $P$ works?}}}
\textbf{Does a small $P$ works?}
We further manually set $P$ as a small value (e.g., $\{2,4,6,8,16\}$).
From the left part (i.e., Small $P$) in Figure  \ref{figure:Evaluation of the hyper-parameters} (a)(d)(g), we can find, that although APG still performs better than Base, there exists performance gap between Large $P$ and Small $P$.
It shows the importance of introducing sufficient shared parameters by over parameterization.
% It shows when the shared parameters

\subsection{The effect of different $K$}
In this part, we evaluate the effect of different $K$.
Specifically, we set $P=32$ and the hidden layers of the deep CTR model as $\{256,128,64\}$ for all cases.
Since $K$ is required to be much smaller than $min(N,M)$, we firstly set $K$ to $\{2,4,6,8\}$ respectively.
The results are reported in the left part (i.e., Small $K$) in Figure  \ref{figure:Evaluation of the hyper-parameters} (b)(e)(h).
Note we also provide the results of the Base 
% and APG w/o re-parameterization (denoted as APG ($\bm{W}_i$)) 
for comparison.
We can find that 
(1) APG performs better than Base in all cases which shows the effectiveness of APG;
(2) With the increasing of $K$, APG can also achieve better performance.
It indicates that it is important to give sufficient specific parameters to characterize the custom patterns of different instances.
% (3) We also notice that when $K$ is set to an extremely small value (e.g., $2$ or $4$), APG performs worse than APG ($\bm{W}_i$).
% One of the possible reasons is that adopting too small values of $K$ makes it difficult to capture the custom patterns.
% While APG ($\bm{W}_i$) uses a large custom weight matrix with heavy computation.

% \subsubsection*{\normalfont{\textbf{Does a large $K$ helps?}}}
\textbf{Does a large $K$ helps?}
Similar to the purpose of over parameterization, ignoring the heavy cost of storage and computation, we can also set $K$ to a large value (e.g., $\{16,32,64,128,256,512\}$) to see whether the model can obtain further improvements.
The results are presented in the right part (i.e., Large $K$) in Figure  \ref{figure:Evaluation of the hyper-parameters} (b)(e)(h).
Some observations are summarized as  follows:
(1) When we set a large value of $K$, compared with the small one, APG can further improve the model performance in most cases.
It indicates increasing $K$ to a large value does help to model the custom patterns;
(2) When $K$ is set to extremely large values (e.g., $256$ or $512$), APG only achieves similar performance with the cases where $K \in \{16,32,64,128\}$;
It shows enlarge the specific parameters does not always give a positive contribution and it is wiser to set a suitable value of $K$.

\subsection{The effect of the different number of layers}
Here, we increase the number of the MLP layers in APG to evaluate the performance.
We also report the results of Base for comparison.
The results are depicted in Figure  \ref{figure:Evaluation of the hyper-parameters} (c)(f)(i).
Some observations are summarized as follows (1) Compared with Base, APG with a different number of layers can always achieve significant improvement;
(2) Compared with the performance of a single layer, increasing the number of layers can make a slight improvement.

\section{Performance in Industrial Sponsored Search System}
\label{sec:Evaluation on the industrial application}
\begin{table}[t]
% \begin{wraptable}{r}{4cm}
 \centering
 % \vspace{-1em}
\caption{The results of the severing efficiency. 
v1 refers to the basic model of APG.}
% \resizebox{4cm}{!}{
\begin{tabular}{l|c|c|c}
\toprule
    & \textbf{Base}  & \textbf{APG} & \textbf{v1} \\ 
    \midrule
RT(ms) & 14.5 &14.8   &   57.1   \\ \midrule
PVR(\%) & 0 & -0.01 & -33.2 \\ \midrule
Memory (M) & 138.03 &16.14 & 4812.7 \\
 \bottomrule
\end{tabular}
% \vspace{-2em}
\label{table:The results of the evaluation of approximate-based strategy. Note RP refers to the Re-Parameterization}
% \end{table}
% }
\end{table} 
Here, we show the performance of APG on industrial applications.
Specifically, we firstly train APG with the industrial dataset and then develop it in the sponsored search system.
Since December 2021, APG is developed and served as the main traffic of our system.
% The overall improvement of APG is summarized in Table \ref{table:The industrial results}.

\begin{figure}
  \centering
    % \subfigure[AUC Gains with different User Group]
     % \subfigure{
    \includegraphics[trim=10 10 10 0, clip,width= .45\linewidth]{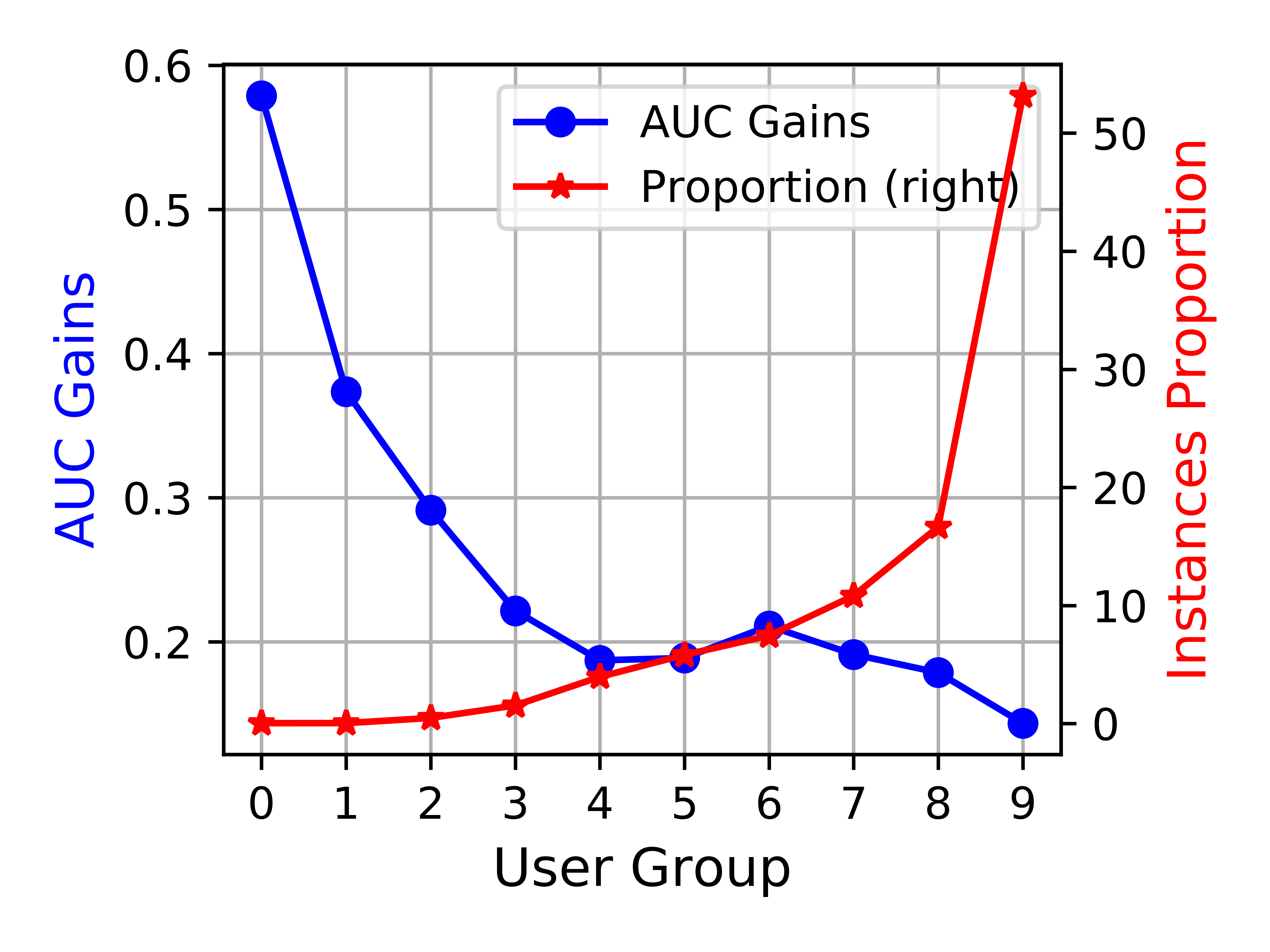}
    % \label{fig:subfigure2}
  % }
      % \subfigure[CTR Gains with different User Group]
     % \subfigure{
    \includegraphics[trim=10 10 10 0, clip,width= .45\linewidth]{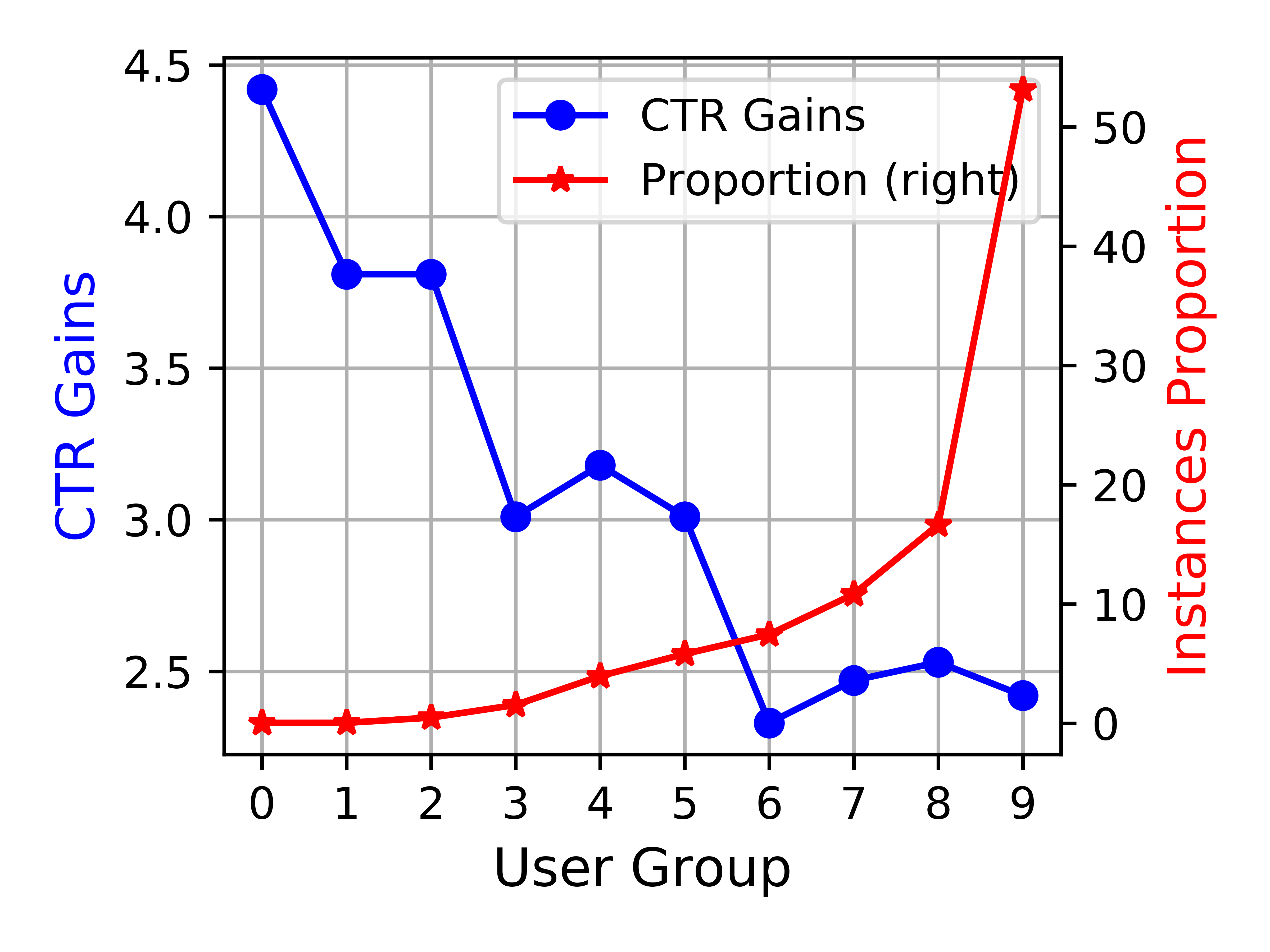}
    % \label{fig:subfigure2}
  % }
\vspace{-1em}
\caption{The AUC (\textbf{Left}) and CTR (\textbf{Right}) gains in different user groups.
 % Note the frequency of user group is increased from group 0 to group 9.
 }
\hspace{1em}
\label{figure:The improvement in different user groups.}
\end{figure}

% \subsubsection*{\normalfont{\textbf{Overall Gains.}}}
\textbf{Overall Gains.}
% \subsubsection{Results on industrial application}
% \label{sec:Results on industrial application}
% The results are reported in Table \ref{table:The industrial results}.
Compared with the online model, it achieves 0.2\% gains in AUC.
During the online A/B test, we observe a 3\% CTR gain and 1\% RPM (Revenue Per Mile) gain respectively.
Note this is a significant improvement in the industrial sponsored search system.
 % which shows the great effectiveness of APG.

% \subsubsection*{\normalfont{\textbf{Severing Efficiency.}}}
\textbf{Severing Efficiency.}
We further evaluate the severing efficiency of APG.
The average Response Time (RT) and Page View Rate (PVR) of the model inference online are evaluated when a user search queries.
We also present the memory cost of the hidden layers in each model.
Not PVR is influenced by the request timeout.
% We also provide the performance of the base model and the version of APG w/o Re-Parameterization. 
The results are reported in Table \ref{table:The results of the evaluation of approximate-based strategy. Note RP refers to the Re-Parameterization}.
Considering the memory efficiency, the memory cost of APG is 8$\times$ smaller than Base.
Considering the time efficiency, APG does not achieve much improvement and has similar RT and PVR compared to Base.
The reason is that, in online serving, the inference time may not have a direct positive relation with GFlops (or theoretical complexity) since the calculation is not always the bottleneck due to the powerful distributed environment and other factors (e.g., I/O, cpu-gpu communication and etc.) also have a great impact on the inference time.
In addition, compared with v1, APG achieves a significant improvement in time cost and memory usage, which demonstrates the efficiency of the proposed extensions in APG.
% We can find APG achieves similar RT and PVR compared
% Some observations can be summarized as follows:
% (1) Compared with Base, APG achieves similar performance in RT and PVR.
% It indicates APG is a lightweight paradigm that does not need too much time cost to allow the model parameters sensitive to different instances;
% (2) v1 needs 3$\times$ times in RT and the PVR also drops 30\%.
% It shows the importance of introducing Re-Parameterization.
% Otherwise, such a model with heavy computation can never be tolerated in real applications.

% \subsubsection*{\normalfont{\textbf{The effect on low-frequency users.}}}
\textbf{The effect on low-frequency users.}
As described in Introduction Section, adding the specific parameters can capture the custom patterns for different instances, especially for the long-tailed instances, since without the specific parameters the model may be easily dominated by the hot instances.
Thus, we detailedly analyze the impact on the different instances when considering the specific parameters.
Specifically, we divide the users into 10 groups by their frequency.
The frequency is increased from group 0 to group 9 and the number of users is set the same in different groups.
Then we evaluate the improvement (including AUC and CTR) of different groups respectively.
The results are reported in Figure \ref{figure:The improvement in different user groups.}.
We can find that 
(1) Since group 9 refers to the users with the highest frequency, although it has only 10\% users, this group produces more than 50\% instances (see the red line in Figure \ref{figure:The improvement in different user groups.}.);
(2) The specific parameters contribute more for low-frequency users since it achieves more gains of AUC and CTR in low-frequency users (e.g., group 0).
It demonstrates that specific parameters do allow low-frequency instances to better represent their features, leading to better performance.

\end{document}